\documentclass[showpacs,twocolumn,superscriptaddress,pre,
preprintnumbers,footinbib,floatfix]{revtex4-1} 

\pdfoutput=1

\usepackage{graphicx}
\usepackage{dcolumn}
\usepackage{bm}
\usepackage{color}
\usepackage{amsmath}
\usepackage{amssymb}

\usepackage{subfigure}

\usepackage{hyperref}
\hypersetup{
  colorlinks,
  citecolor=red,
  filecolor=black,
  linkcolor=blue,
  urlcolor=magenta
}

\usepackage{mathptmx}

\newcommand{\bpsi}{\boldsymbol{\psi}}
\newcommand{\XXiflow}{\Xi_\mathrm{flow}}
\newcommand{\XXidisc}{\Xi_\mathrm{disc}}

\newcommand{\xiflow}{\xi_\mathrm{flow}}
\newcommand{\xidisc}{\xi_\mathrm{disc}}

\newcommand{\rr}{\mathbf{r}}
\newcommand{\vv}{\mathbf{v}}

\newcommand{\fref}[1]{Fig.~\ref{#1}}
\renewcommand{\ln}{\log}

\newcommand{\taumft}{\tau_\mathrm{MFT}}
\newcommand{\tauex}{\tau_\mathrm{R}}
\newcommand{\tauball}{\tau_\mathrm{B}}

\begin{document}

\title{Measuring logarithmic corrections to normal diffusion in infinite-horizon billiards}

\author{Giampaolo Cristadoro}
\email{giampaolo.cristadoro@unibo.it}
\affiliation{Dipartimento di Matematica, Universit\`a di Bologna, 
Piazza di Porta S. Donato 5, 40126 Bologna, Italy}
\author{Thomas Gilbert}
\email{thomas.gilbert@ulb.ac.be}
\affiliation{Center for Nonlinear Phenomena and Complex Systems,
  Universit\'e Libre  de Bruxelles, C.~P.~231, Campus Plaine, B-1050
  Brussels, Belgium}
\author{Marco Lenci}§
\email{marco.lenci@unibo.it}
\affiliation{Dipartimento di Matematica, Universit\`a di Bologna, 
Piazza di Porta S. Donato 5, 40126 Bologna, Italy}
\affiliation{Istituto Nazionale di Fisica Nucleare, Sezione di
  Bologna, Via Irnerio 46, 40126 Bologna, Italy}
\author{David P.~Sanders}
\email{dpsanders@ciencias.unam.mx}
\affiliation{Departamento de F\'isica, Facultad de Ciencias, Universidad
Nacional Aut\'onoma de M\'exico,  Ciudad Universitaria,
04510 M\'exico D.F.,
 Mexico}
\date{\today}

\begin{abstract}
  We perform numerical measurements of the moments of the position of
  a tracer particle in a two-dimensional periodic billiard model
  (Lorentz gas) with infinite corridors. This model is known to
  exhibit  a weak form of  super-diffusion, in the sense that there is
  a logarithmic correction to the linear growth in time of the
  mean-squared displacement. We show numerically that this expected
  asymptotic behavior is easily overwhelmed by the subleading linear
  growth throughout the time-range accessible to numerical
    simulations. We compare our simulations to the known analytical
  results for the variance of the anomalously-rescaled limiting normal
  distributions.
\end{abstract}


\maketitle


Billiard models are among the simplest dynamical systems, and have
proven suited to model  problems pertaining to a variety of fields,
from experimental to mathematical physics. Within the framework of
statistical mechanics and nonlinear dynamics, the Lorentz gas, which
consists of a point-like particle moving freely and bouncing
elastically off a set of fixed circular scatterers, has served as a
paradigm to study transport properties of light particles among
heavier ones \cite{Chernov:2006p683, Gaspard:1998book,
  Szasz:2000book, Cvitanovic:2004p284, 
  Dettmann:2000inSzasz, Gaspard:2003p298}. 
Its lasting popularity is due, in particular, to the fact that it
allows to choose different geometries  (disordered or periodic
arrangements of the scatterers), and to identify regimes of both
normal and anomalous transport. 

In two dimensions, when the geometry is periodic and chosen in such a
way that the distance between any two successive collisions  is
bounded above (the so-called \emph{finite-horizon condition}), it is
known that the transport is normal, which is to say that the  distribution
of the the displacement vector is asymptotically Gaussian, with
a variance growing linearly in time \cite{Bunimovich:1981p479,
 Bunimovich:1991p47, Chernov:2006v122p1061}. Indeed, a large body of
research on such dispersing \emph{Sinai billiards} has produced a
number of rigorous results about the statistical and transport
properties of the periodic Lorentz gas \cite{Chernov:2006p683}, among
which are the exponential decay of correlations for periodic observables
\cite{Young:1998p136},  the central limit theorem and invariance
principle, i.e., convergence to a  Wiener process
\cite{Bunimovich:1981p479, Bunimovich:1991p47, 
  Chernov:2006v122p1061}, and  recurrence  \cite{Schmidt:1998v327p837, 
  Conze:1999v19p1233}. The same types of results would give recurrence
for the typical \emph{aperiodic} gas as well \cite{Lenci:2003v23p869,
  Lenci:2006v26p799, cristadoro:2010recurrence}, but proving normal
diffusion in that case remains an open problem \cite{Chernov:2006v2p1679}. 
See also the recent survey \cite{dettmann:1402.7010}.

In this article, we are concerned with infinite-horizon periodic
Lorentz gases, i.e., such that point particles can move arbitrarily
far through regions devoid of obstacles. We refer to these regions as
\emph{corridors}, following Ref.~\cite{1992JSP....66..315B}; they are
elsewhere termed gaps  \cite{Sanders:2008p453},  horizons
\cite{Dettmann:2011p18216}, and free planes \cite{Nandori:2012arXiv1210}. 

The presence of such regions leads to qualitatively different
transport than the finite-horizon case, with a weak 
form of super-diffusion, in the sense that there is a
logarithmic correction to the linear growth in time of the
mean-squared displacement  \cite{Zacherl:1986p7768}. The  diffusion coefficient thus
diverges, as initially suggested by Friedman and Martin
\cite{Friedman1984p23, Friedman:1988v30p219}.  There and elsewhere
\cite{GarridoGallavottiBilliardCorrelationFnsJSP1994,
  Matsuoka:1997p776}, numerical studies of this logarithmic correction
often focused on velocity autocorrelation functions, which are
expected to decay like $1/t$  
\cite{Dahlqvist:1996p16292, Melbourne:2009v98p163}.   

Bleher \cite{1992JSP....66..315B} gave a semi-rigorous discussion of 
super-diffusion in the infinite-horizon Lorentz gas. A number of proofs
were subsequently obtained for the discrete-time collision map
by Sz\'asz and Varj\'u \cite{Szasz:2007v129p59},  including a local limit law to a
normal distribution, as well as recurrence and ergodicity in the full
space. Techniques there were based on work of B\'alint and Gou\"ezel
\cite{Balint:2006p18224} for the stadium billiard, which also has long
segments of trajectories without a collision on a curved boundary and
a normal distribution with a non-standard  limit law. 
The extension to the continuous-time dynamics was subsequently proved
by Chernov and Dolgopyat \cite{Dolgopyat:2009p16456}, who, in addition, proved 
the weak invariance principle in this case. 

While rigorous results are of major theoretical importance, they have
thus far not been complemented by convincing numerical measurements of
the  logarithmic correction to the linear growth in time of the
mean-squared displacement, which has proven difficult to characterize
\cite{Zacherl:1986p7768,  
  GarridoGallavottiBilliardCorrelationFnsJSP1994}. Though some authors
have reported numerical evidence of this growth
\cite{Courbage:2008p454},  the only attempt known to us to confront
results with known analytic formulae for the asymptotic behavior of the
mean-squared displacements  
\cite{1992JSP....66..315B, Szasz:2007v129p59, Dolgopyat:2009p16456}
was met with limited success \cite{Dettmann:2011p18216}. 

Indeed, a main problem lies in trying to identify only the logarithmic
divergence of  the mean-squared displacement $\langle \| \rr(t) -
\rr(0) \|^{2} \rangle$, while ignoring other relevant terms in its time
dependence; see  Sec.~\ref{sec:model} for precise definitions.  As
stated by Bleher \cite[Eq.~(1.9)]{1992JSP....66..315B},
when $t \to \infty$, the \emph{finite-time diffusion coefficient},
$D(t)$, has the asymptotic behavior 
\begin{equation}
  D(t) \equiv \frac{\langle \| \rr(t) - \rr(0) \|^{2} \rangle}{4t} \sim \ln t;
  \label{eq:ftimediffcoeff}
\end{equation}
see also references \cite{Dahlqvist:1996p16292,
    Dahlqvist:1996p16294}.
However, this asymptotic behavior is attained when $\ln t \gg
1$, which is numerically unattainable; cf.\ discussion below. 
In the pre-asymptotic regime, other terms must also be taken into
account on the right-hand side of this expression, most notably a
constant term, which may actually turn out 
to be the largest contribution when $t$ is large but $\ln t$ is not. 
Failing to do so, for example, by considering the mean-squared
displacement as a function of $t \ln t$ \cite{Courbage:2008p454},
masks the relative contributions of the two terms, and 
hence does not allow to accurately measure either of them.  

Rather, it is necessary to consider the finite-time diffusion
coefficient~\eqref{eq:ftimediffcoeff} as an asymptotically
affine function of $\ln t$, taking into account both the intercept
and the slope, as was previously applied by one of the present authors
in other super-diffusive billiard models \cite{Sanders:2006p452,
  Sanders:2008p453}. When $t \to \infty$, the slope is the
second moment of the rescaled process and thus characterizes the
strength of this type of super-diffusion. The physical interpretation of
the intercept may, on the other hand, not always be clear, for
example, when this quantity is negative. However, for a system which
exhibits normal diffusion, this quantity obviously reduces to the
standard diffusion coefficient. By extension, at least so long as
the slope is small compared to the intercept, we will think of the
intercept as accounting for a diffusive component of the
process, coexisting with the anomalous diffusion. 

In this paper, we report numerical measurements of these 
quantities for continuous-time dynamics in two-dimensional periodic
Lorentz gases with infinite horizon,  comparing them  to analytical 
asymptotic results. 
We  point out several difficulties  arising in the numerical analysis.
The main one, is that large fluctuations underlie the super-diffusive
regime, and these require a very large number of initial conditions for the
logarithmic divergences to be observed with sufficient precision. 
A typical trajectory exhibits long paths free of collisions, whose
frequency of occurrence decays with the cube of their lengths
\cite{1992JSP....66..315B}. We refer to these free paths as ballistic
segments, i.e., segments of a trajectory separating two successive
collisions with obstacles. Although they may be rare, long ballistic
segments contribute significantly to the mean-squared displacement
measured over the corresponding time scale.

This observation points to a second, often underestimated problem,
which is that the  integration time should be neither too short, nor
too long; on the one hand there are long transients before the
asymptotic behavior sets in, so that integration times must not be
too short; on the other hand, long integration times require
averaging over prohibitively large number of trajectories to achieve a
proper sampling of ballistic segments. 

The paper is organized as follows. In Sec.~\ref{sec:model}, we define
the infinite-horizon periodic Lorentz gas and identify the single relevant
parameter. The statistical properties of trajectories are discussed in
Sec.~\ref{sec:convergence}, where we obtain the asymptotic
distribution of the anomalously-rescaled displacement vector. The
variance of this distribution is given in
Sec.~\ref{sec:variance}. In Sec.~\ref{sec.numerics}, we present numerical
computations of the first two moments of the rescaled displacement
vector in the infinite-horizon Lorentz gas and compare them to 
asymptotic results. 
Conclusions are drawn in Sec.~\ref{sec:conclusion}. 

\section{Infinite-horizon Lorentz gas
\label{sec:model}
}

We study the  periodic Lorentz gas on a two-dimensional square
lattice, which is the simplest billiard model with infinite
horizon. This is constructed starting from  a Sinai billiard with a
single circular scatterer of radius $0 < \rho < \ell/2$ at the center
of a square cell with side length $\ell$, taken to be $\ell \equiv 1$,
and periodic boundary conditions. Unfolding this onto the whole of
$\mathbb{R}^{2}$ produces a square lattice of obstacles (or scatterers),
the periodic Lorentz gas. We refer to the contour of the obstacles as
the boundary of the billiard table. 

\begin{figure}[tb]
  \centering
  \includegraphics[width=0.35\textwidth]{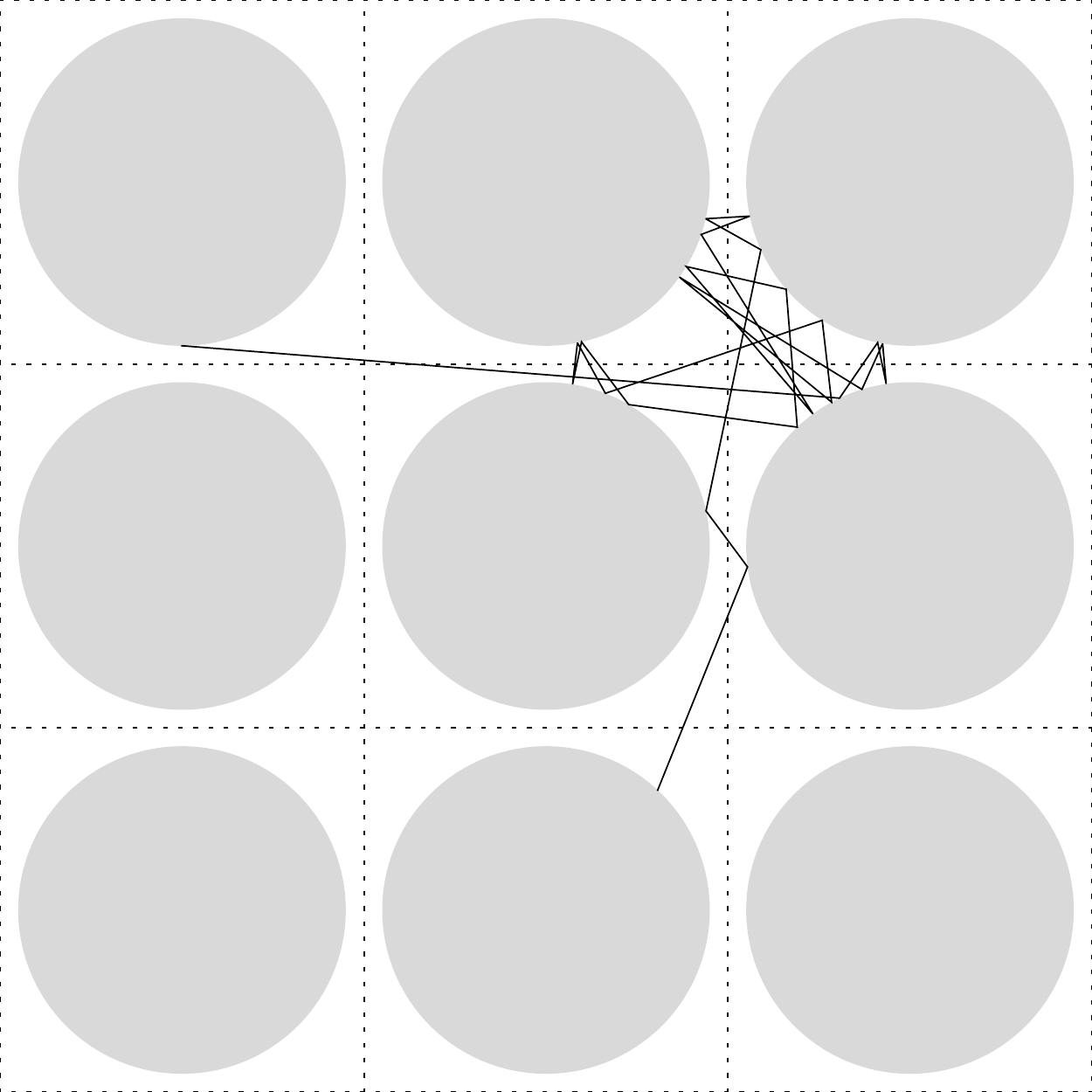}
  \caption{A trajectory in the infinite-horizon periodic Lorentz
    gas. 
  }   
  \label{fig:trajectory}
\end{figure}

The dynamics consist of point particles with unit speed that move
freely between the obstacles until they collide with one of them. They
then undergo an elastic collision, i.e., such that the angle of
reflection is equal to the angle of incidence, and proceed to the next
collision. Figure~\ref{fig:trajectory} shows an example trajectory in
this billiard table. The infinite horizon is synonymous with the
existence of ballistic trajectories, such as horizontal or vertical
trajectories along the corridors spreading about the dotted lines in 
\fref{fig:trajectory}. New corridors appear as the model's parameter
$\rho$  decreases; see the discussion in Sec.~\ref{sec:variance}. 

As is standard in billiard models, the dynamics may be studied either
in discrete time or continuous time. 
The former is referred to as the billiard map, the latter as the
billiard flow. 
The billiard map is restricted to the boundary of the billiard, and
maps points with outgoing velocity from one collision to the next.
The position at the $n$th collision will be denoted $\rr_{n}$ and its
velocity $\vv_{n}$.  Points under the billiard flow have position
$\rr(t)$ and velocity $\vv(t)$ at continuous time $t$. Denoting the
time of the $n$th collision by $t_{n}$, we thus have $\rr(t_{n}) =
\rr_{n}$, while, for $t_{n} < t < t_{n+1}$, $\rr(t)$ is a point on the
straight line joining $\rr_{n}$ and $\rr_{n+1}$, such that $\rr(t) =
\rr_{n} + (t - t_{n})\vv_{n}$.  Correspondingly, the velocity $\vv(t)$
remains unchanged until the next collision, $\vv(t) = \vv_n$.
Due to the existence of open corridors, particles may
propagate arbitrarily far without collision, so that the ballistic
segments that connect successive collisions $\rr_{n}$ and $\rr_{n+1}$
are unbounded.  

\section{Convergence to asymptotic behavior
  \label{sec:convergence}
}

We are interested in asymptotic transport properties, i.e., in the
distribution of the displacement vector  $\rr(t) - \rr(0)$ in the
limit $t \to \infty$.  In the remainder we shorten the notation and
denote the displacement vector  simply by $\rr(t)$ as  no confusion
will arise.  
The modulus of this quantity will be denoted by $r(t)$.

\subsection{Convergence in distribution
  \label{sec:convdistrib}
} 

For the discrete-time dynamics, it has been proved
\cite{Szasz:2007v129p59, Dolgopyat:2009p16456} 
that the displacement vector distribution with anomalous rescaling
converges in distribution to a centered normal distribution: that is,
as $n \to \infty$, 
\begin{equation}
  \label{eq:ct-weak-conv}
  \frac{ \rr_n } {\sqrt{n \ln{n}}} \to_\mathrm{d} \mathcal{N}(0,\XXidisc),
\end{equation}
which means that the probability that the quantity in the left-hand
side lies in a regular set $K$ converges to the probability that a
normally-distributed random variable with mean $0$ and variance matrix
$\XXidisc$ lies in $K$.  

The covariance matrix $\XXidisc$ is a multiple of the identity matrix,
i.e., its entries are given by $(\XXidisc)_{i,j}=\xidisc
\delta_{i,j}$. The  \emph{discrete-time limiting variance} $\xidisc$
is expressed in terms of the geometrical parameters of the model in
Sec.~\ref{sec:variance}. 

The corresponding result for the continuous-time flow was proved in
\cite{1992JSP....66..315B, Dolgopyat:2009p16456}, and states the
following: as $t \to \infty$,  
\begin{equation}
  \label{eq:weak-convergence}
  \frac{ \rr(t)}{\sqrt{t \ln t}} \to_\mathrm{d}
  \mathcal{N}(0,\XXiflow), 
\end{equation}
where $(\XXiflow)_{i,j} =\xiflow \, \delta_{i,j}$ and $\xiflow=\xidisc / \taumft$.
Here, $\taumft$ is the mean free time between collisions, which is
proportional to the available area in the unit cell,  $1 - \pi
\rho^{2}$, divided by the perimeter of the boundary, $2 \pi \rho$
\cite{Chernov:1997p1}:   
\begin{equation}
  \label{eq:mftime}
  \taumft = \frac{1-\pi \rho^{2}}{2 \rho}.
\end{equation}

\subsection{Asymptotic behavior of moments 
  \label{sec.asymp}
}

A standard method to characterize convergence of random variables
numerically is via their moments. It is important to note, however,
that convergence in distribution of a sequence of random variables to
a limiting distribution does not necessarily imply convergence of the
moments of the sequence to the moments of the limiting distribution. 

Indeed, Armstead et al. \cite{2003PhRvE..67b1110A} showed that the
moments have dominant behavior:
\begin{equation}
  \langle r(t)^q \rangle \sim 
  \begin{cases} 
    t^{q/2}, & q < 2, \\
    t, & q = 2, \\    
    t^{q-1}, &q > 2, 
  \end{cases}
  \label{eq:armstead-moments}
\end{equation}
ignoring  logarithmic corrections; see also
Ref.~\cite{2003PhRvL..90x4101A}. A proof of the  result for $q>2$ has 
recently been announced \cite{Melbounre:2012v32p1091,
  Melbourne:private}.

The type of qualitative change in the scaling of the moments seen in
Eq.~\eqref{eq:armstead-moments} has elsewhere been dubbed
\emph{strong  anomalous diffusion} \cite{Castiglione:1999p690}, as opposed to
weak when a single exponent ($\neq 1/2$) characterizes the whole spectrum of
moments. It appears to be typical for anomalous transport arising from
deterministic dynamical systems \cite{2003PhRvL..90x4101A}, as opposed
to the single  scaling of the converging moments for self-similar
stable distributions \cite{gnedenko1968limit}. Such behavior is due to
the fact that the slowly-decaying tail of the displacement vector
distribution  may give no contribution to the  convergence in
distribution of the rescaled variable, while nonetheless playing a
dominant role for sufficiently high moments.

We denote by $M_{q}$ the $q$th moment of the limiting two-dimensional
normal distribution \eqref{eq:weak-convergence}: 
\begin{equation}
  M_{q} 
  \equiv \Gamma \left( 1+\tfrac{q}{2} \right) \, (2
  \xiflow)^{q/2}, 
  \label{eq:normal-moments} 
\end{equation}
where $\Gamma$ is the Gamma function.
If the convergence in \eqref{eq:weak-convergence} were sufficiently
strong, then the $q$th moment  of  the rescaled displacement
vector distribution would converge to $M_{q}$, for all $q$.
In fact, however, the weak convergence~\eqref{eq:weak-convergence}
implies this convergence of the moments  
only for $q < 2$ \cite[Sec.~3.2]{Knight:2000MathStat}:  
\begin{equation}
  \left \langle \left [ \frac{r(t)}{\sqrt{t \ln t}} \right ] ^{q}
  \right \rangle \to   M_{q} \qquad (q < 2), 
  \label{eq:asympmoment}
\end{equation}
when $t \to \infty$.
For $q > 2$ this does not hold, and the asymptotic values of
the $q$th moments of the rescaled displacement vector distribution
diverge, as follows from \eqref{eq:armstead-moments}.  

The case $q=2$ requires special consideration. If
Eq.~\eqref{eq:asympmoment} applied in this case, we would have
the asymptotic behavior 
$\langle r(t)^{2} \rangle/(t \ln t) \sim 2 \xiflow$.
However, this is incorrect: it has recently been discovered that in
fact an extra factor of $2$ appears,  so that the correct asymptotic
behavior is 
\begin{equation}
  \frac{1}{2t} \left \langle r(t)^{2} \right \rangle  \sim 2 \xiflow \ln t.
  \label{eq:moment2}
\end{equation}
An explanation of this phenomenon was given in
Ref.~\cite{Dettmann:2011p18216}, and a  proof
has been announced \cite{Chernovetal:private}.
A similar result appears also in a different setting; see
Ref.~\cite{Balint:2011p18227}, where a rigorous argument is available   
for a related billiard model with cusps. 
This surprising behavior is due to the fact that
the contribution to the second moment \eqref{eq:moment2} of
collisionless orbits is equal to that coming from the central part of
the distribution, while playing no role in the weak convergence to a
normal distribution in Eq.~\eqref{eq:weak-convergence}. 

We note that there has been recent interest in the extension
of these results to higher-dimensional Lorentz gases
\cite{Sanders:2008p453, Dettmann:2011p18216, Nandori:2012arXiv1210},
where additional effects come into play.

\section{Corridors and variance of limiting distribution 
  \label{sec:variance}
}

Before turning to numerical measurements of the asymptotic behaviors 
\eqref{eq:asympmoment} and \eqref{eq:moment2} in the next section, we 
consider the computation of the variance of the limiting distribution
\eqref{eq:weak-convergence}.

As proved in \cite{Dolgopyat:2009p16456}, the general expression for the discrete-time
limiting covariance matrix $\XXidisc$ is  
\begin{equation}   
  (\XXidisc)_{i,j} = \frac{c_{\nu}}{2}
  \sum_x \frac{w_x^2 \, \psi_i(x) \psi_j(x)}{|\bpsi(x)|},
  \label{eq:variance}
\end{equation}
where the sum runs over all fixed points $x$ of the collision map on the unit
cell, of which there are four for each corridor. Here  $w_x$ is the width of
the corresponding corridor and $\bpsi(x)$ is the vector of Cartesian
components $\psi_i(x)$ ($i = 1,2$) parallel to the
corridor, giving the translation in 
configuration space described by the action of the map on $x$ 
\footnote{In other words, if $x = (\rr,\vv)$
  denotes the phase-space coordinates at a grazing collision point 
  $\rr$, i.e.~such that $\vv$ is tangent to 
  the scatterer, $x$ is mapped to a point $x_1= (\rr_1,\vv)$ by the
  collision map whose velocity component remains unchanged.
  The vector connecting the two successive positions
  is $\bpsi(x) = \rr_1 - \rr$.},
and 
$c_{\nu} = 1/(4 \pi \rho)$ 
is a normalizing constant.
It  follows from the symmetries of the system that the
non-diagonal elements of the covariance matrix vanish.

Given the parameter value $\rho$, Eq.~\eqref{eq:variance}, along with an
enumeration of all the fixed points of the collision map, allows one
to write the expression for the discrete-time limiting variance
$\xidisc$. By including the mean free time \eqref{eq:mftime}, the corresponding
expression for the continuous-time limiting variance $\xiflow$ may then
be obtained.

\begin{figure}[t]
  \centering
  \null\hfill
  \subfigure[~Type $(0,1)$ ($\rho = 0.4$).]{
    \includegraphics[width=0.145\textwidth]{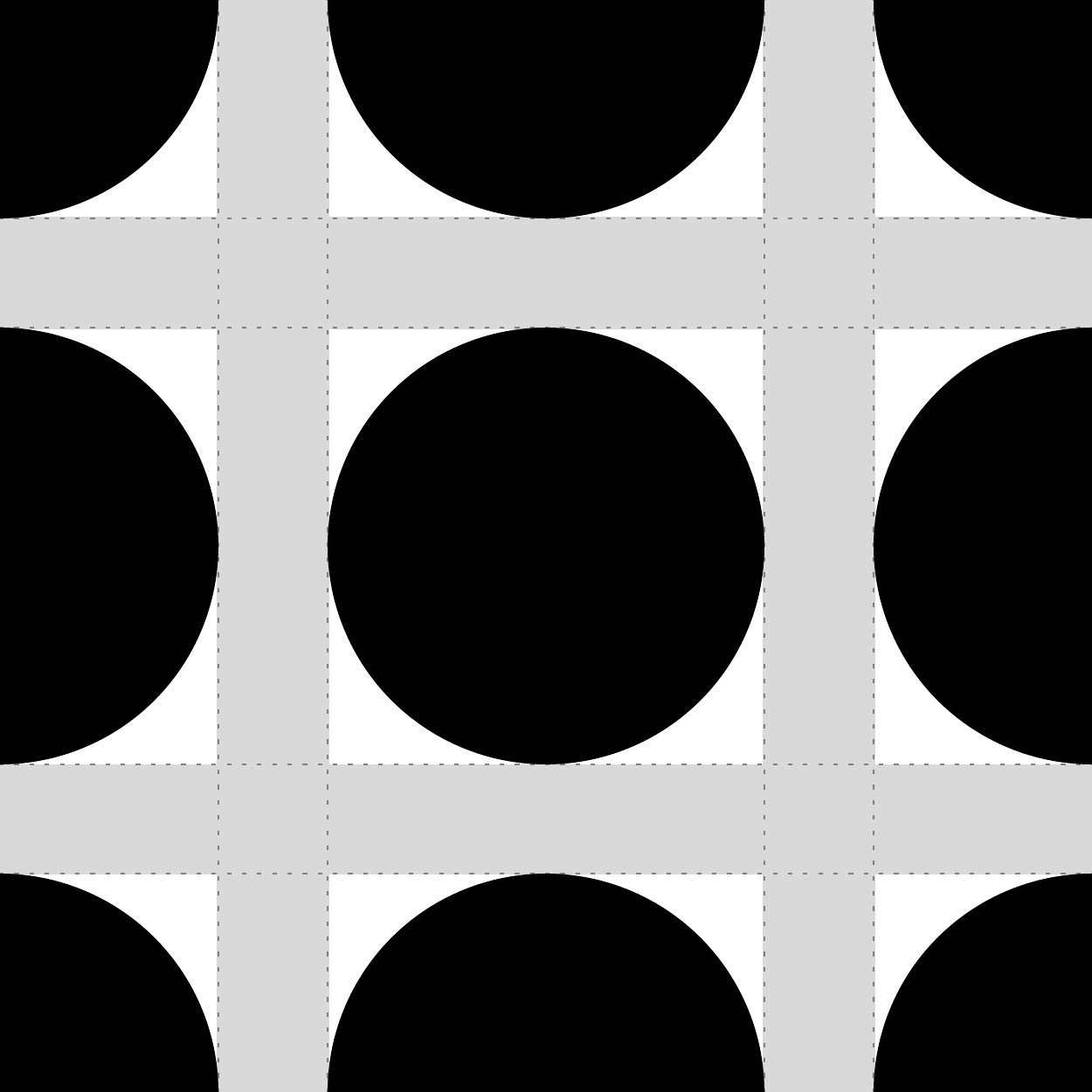}
  }
  \hfill
  \subfigure[~Type $(1,1)$ ($\rho = 0.3$).]{ 
    \includegraphics[width=0.145\textwidth]{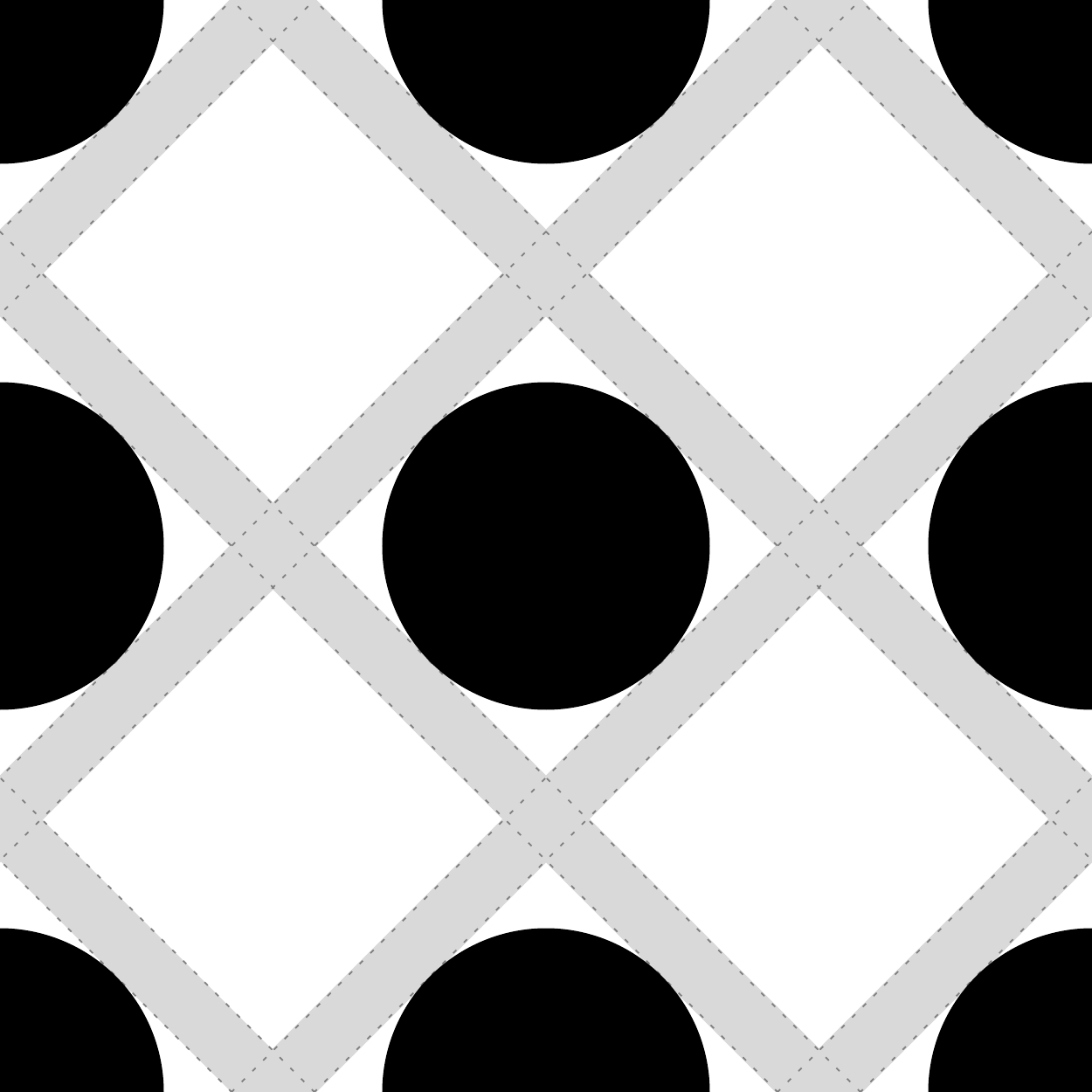}
  }
  \hfill
  \subfigure[~Type $(1,2)$ ($\rho = 0.2$).]{
    \includegraphics[width=0.145\textwidth]{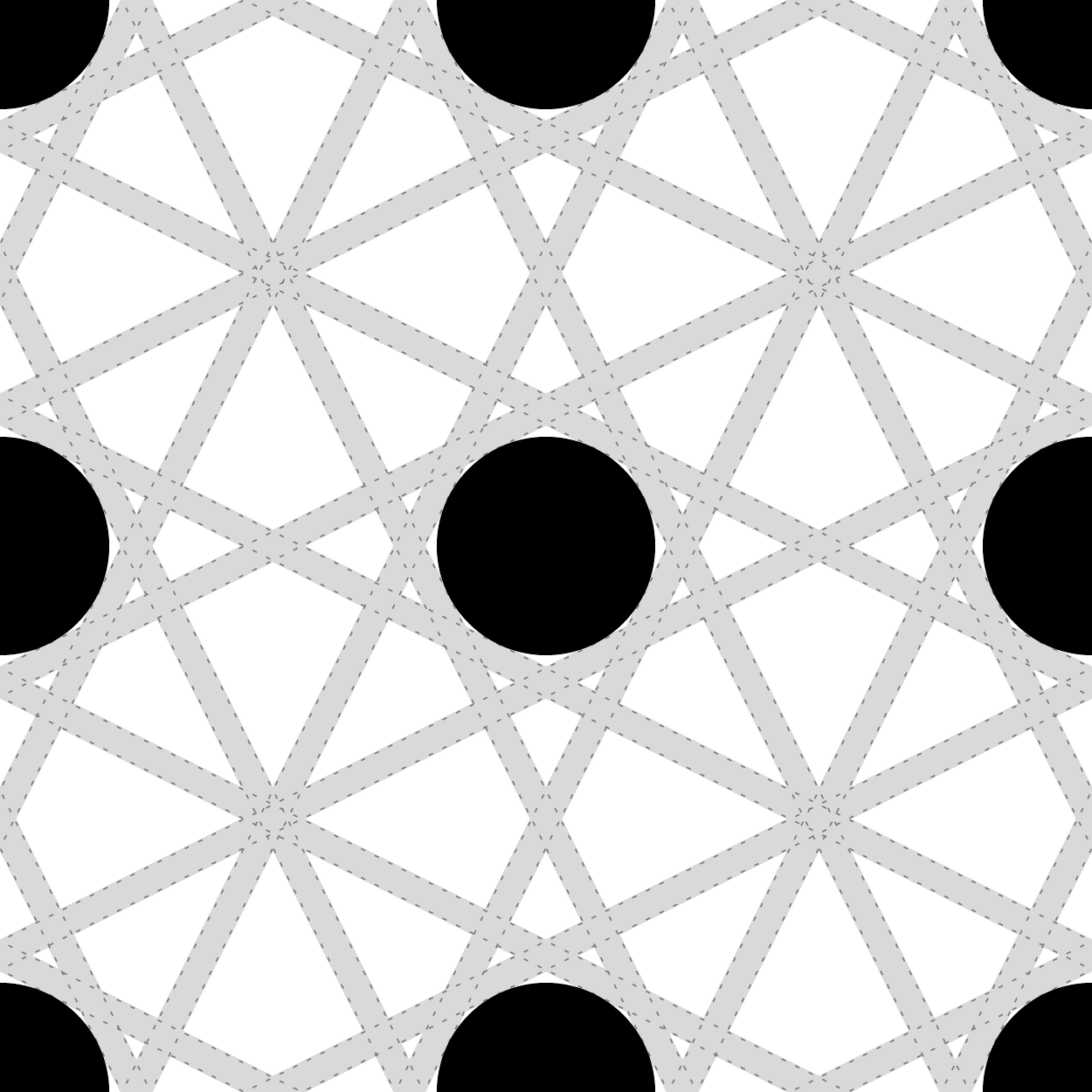}
  }
  \hfill\null
  \caption{Examples of infinite-horizon corridors in the periodic
    Lorentz gas, labelled according to their  type $(m, n)$; see
    text for details. 
    \label{fig:corridors}
  }  
\end{figure}

When $1/(2 \sqrt{2}) \le \rho < 1/2$, the only corridors present are
horizontal and vertical corridors of width $w = 1 - 2\rho$, 
which we refer to as type $(0,1)$ corridors; see \fref{fig:corridors}(a). 
$\xiflow$ then has the only contribution 
\begin{equation}
  \frac{(1 - 2\rho)^2}{\pi(1-\pi\rho^2)}.
  \label{eq:variance10}
\end{equation}

When $1/(2\sqrt{5}) \le \rho < 1/(2\sqrt{2})$, two new corridors, of type
$(1,1)$, open up, along the vectors $(1, \pm 1)$, with width $w =
1/\sqrt{2} - 2\rho$; see \fref{fig:corridors}(b).  
Their contribution to $\xiflow$ is
\begin{equation}  
  \label{eq:variance11}
  \frac{1}{\pi(1-\pi\rho^2)}
  \sqrt 2 \Big(\frac{1}{\sqrt{2}} 
  - 2\rho\Big)^2.
\end{equation}

Additional corridors keep appearing as $\rho$ decreases. By symmetry,
they all occur in quadruplets. For instance, the type $(1,2)$
corridors, which appear when $\rho < 1/(2\sqrt{5})$, are shown in
\fref{fig:corridors}(c);  they point along the vectors $(1, \pm 2)$
and $(2, \pm 1)$ and have width  $w = 1/\sqrt{5} - 2\rho$. 

The general expression of the limiting variance is
\begin{align}
  \xiflow & = \frac{1}{\pi(1-\pi\rho^2)}
  \bigg[(1 - 2\rho)^2 + \sqrt 2 \Upsilon \Big(\frac{1}{\sqrt{2}} 
  - 2\rho\Big)
  \label{eq:variancetotal}\\
  & \quad + 2
  \sum_{m=1}^{\infty}
  \sum_{\substack{n = m+1: \\ \mathrm{gcd}(m,n)=1 }}^{\infty}
  \sqrt{m^2+n^2} \, \,
  \Upsilon\Big(\frac{1}{\sqrt{m^2+n^2}} - 2 \rho\Big)
  \bigg], 
  \nonumber
\end{align}
where $\Upsilon(x) = x^2$ if $x>0$, and $0$ otherwise and
$\mathrm{gcd}(m,n)$ denotes the greatest common divisor of $m$ and
$n$;  the sum thus runs over all pairs of relatively prime integers
$m$ and $n$ such that $1 \le m < n$. 
The number of contributions to $\xiflow$ 
depends on the radius $\rho > 0$, and is always  finite.
For example, for $\rho = 0.2$, there are three types of corridors
open, depicted in \fref{fig:corridors}. 

\section{Numerical measurements of the moments 
  \label{sec.numerics}
}
We study the  behavior of the moments of the distribution of the
anomalously-rescaled process $\rr(t)/\sqrt{ t \ln t}$, relating our
numerical results to the parameters of its limiting  normal
distribution  such as  $\xiflow$, Eq.~\eqref{eq:variancetotal},
which, for simplicity, we will henceforth refer to as the variance.  

\subsection{Time-dependence of the first and second moments}

\begin{figure*}[tbp]
  \centering
  \null\hfill
  \subfigure[~1st and 2nd moments, $\rho = 0.14$]{
    \includegraphics[width=.31\textwidth]{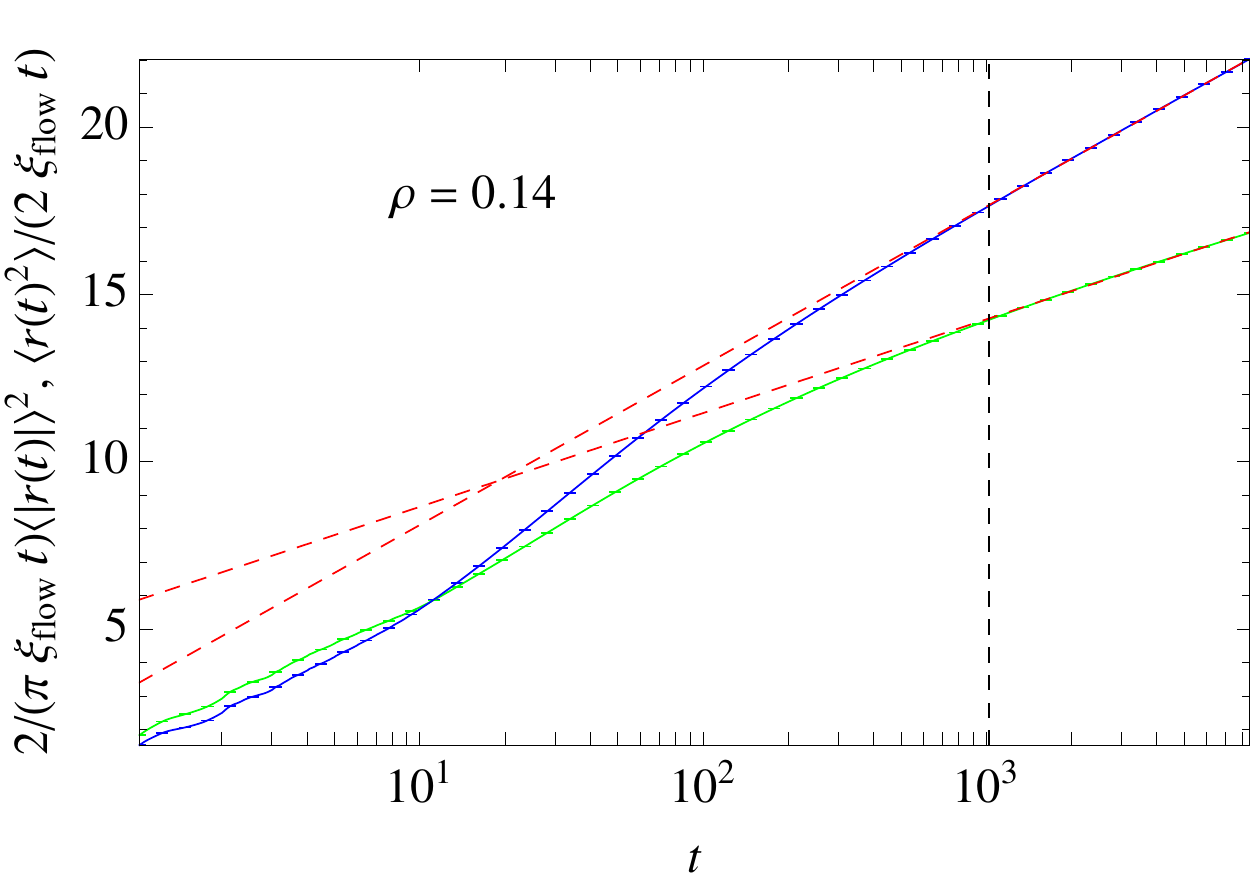}
    \label{fig.moments014}
  }
  \hfill
  \subfigure[~Intercepts, $\rho = 0.14$]{
    \includegraphics[width=.31\textwidth]{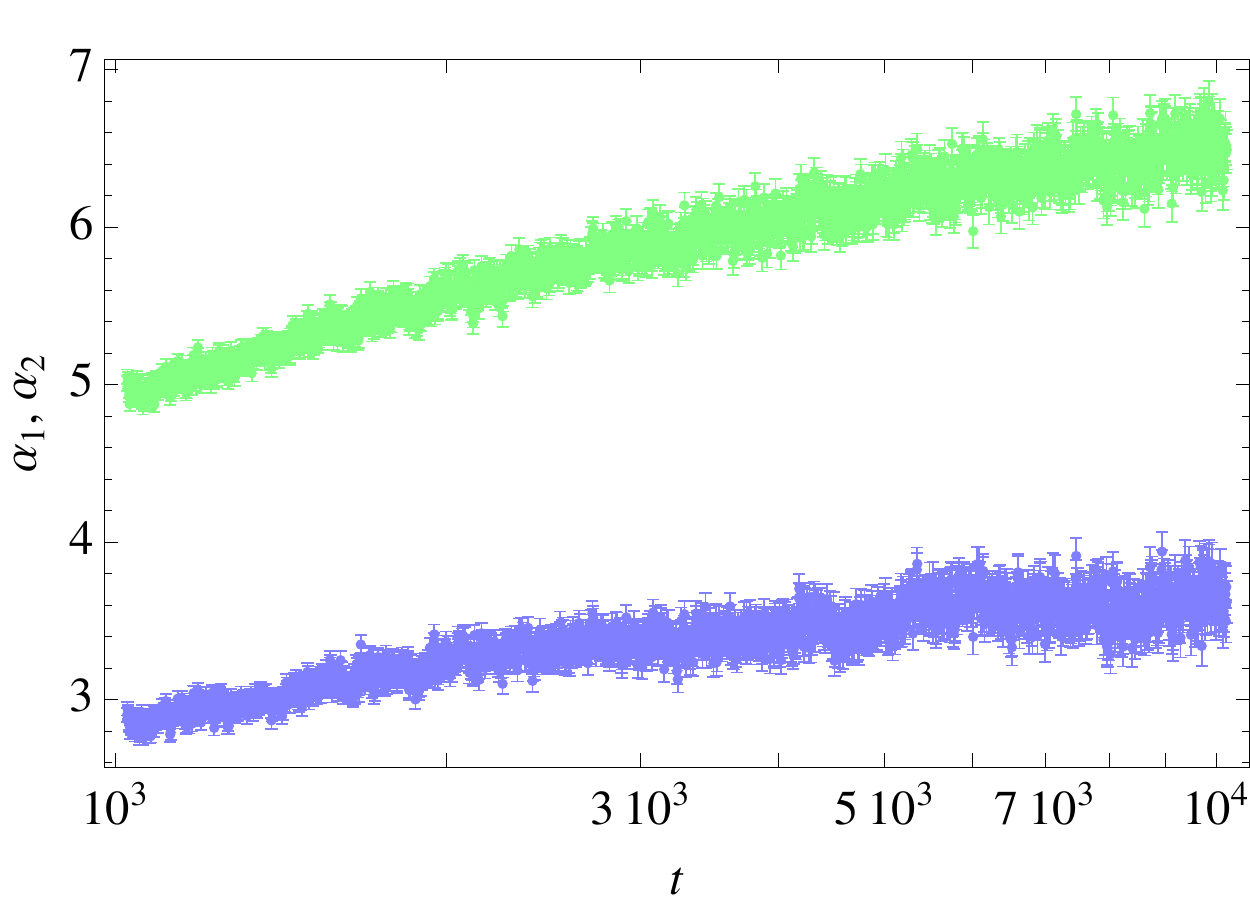}
    \label{fig.alpha014}
  }
  \hfill
  \subfigure[~Slopes, $\rho = 0.14$]{
    \includegraphics[width=.31\textwidth]{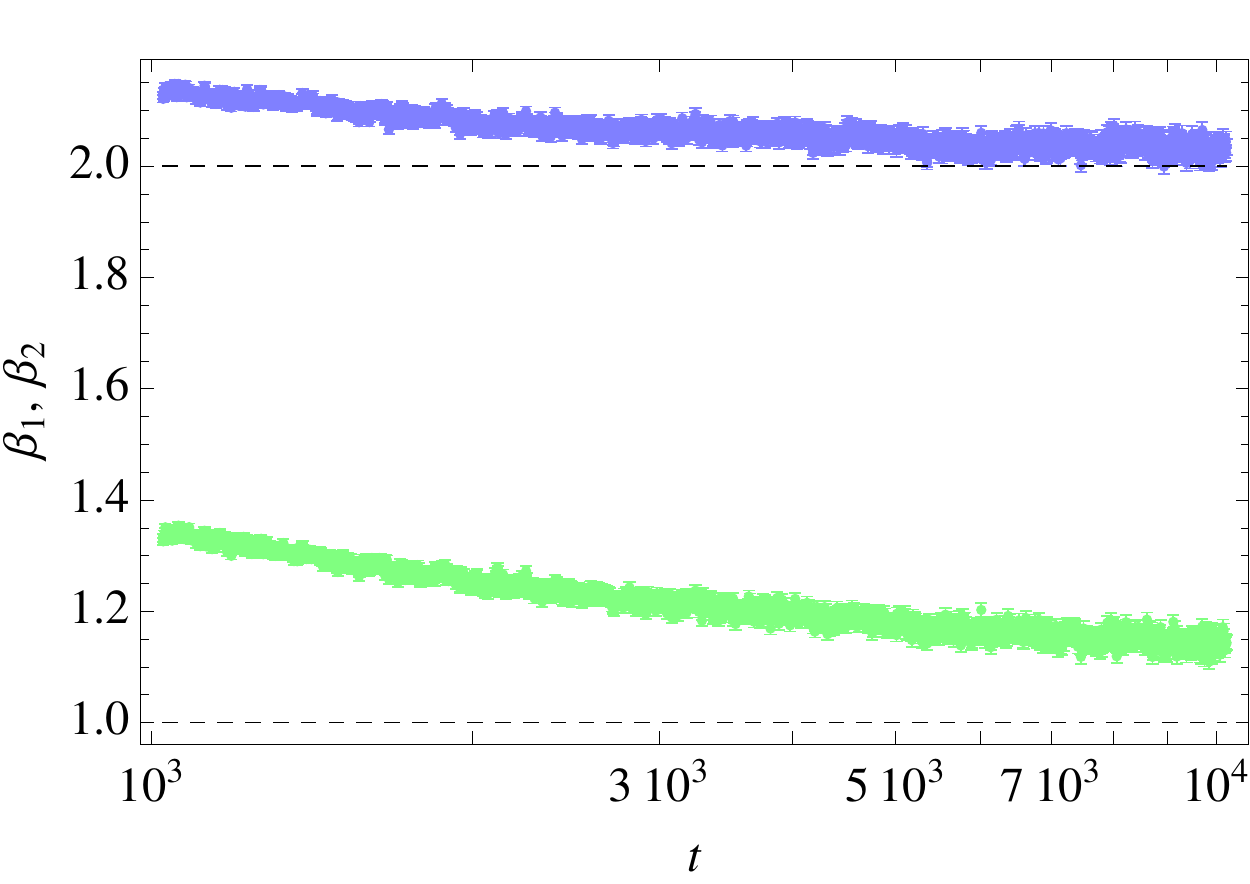}
    \label{fig.beta014}
  }
  \hfill\null
   
  \null\hfill
  \subfigure[~1st and 2nd moments, $\rho = 0.24$]{
    \includegraphics[width=.31\textwidth]{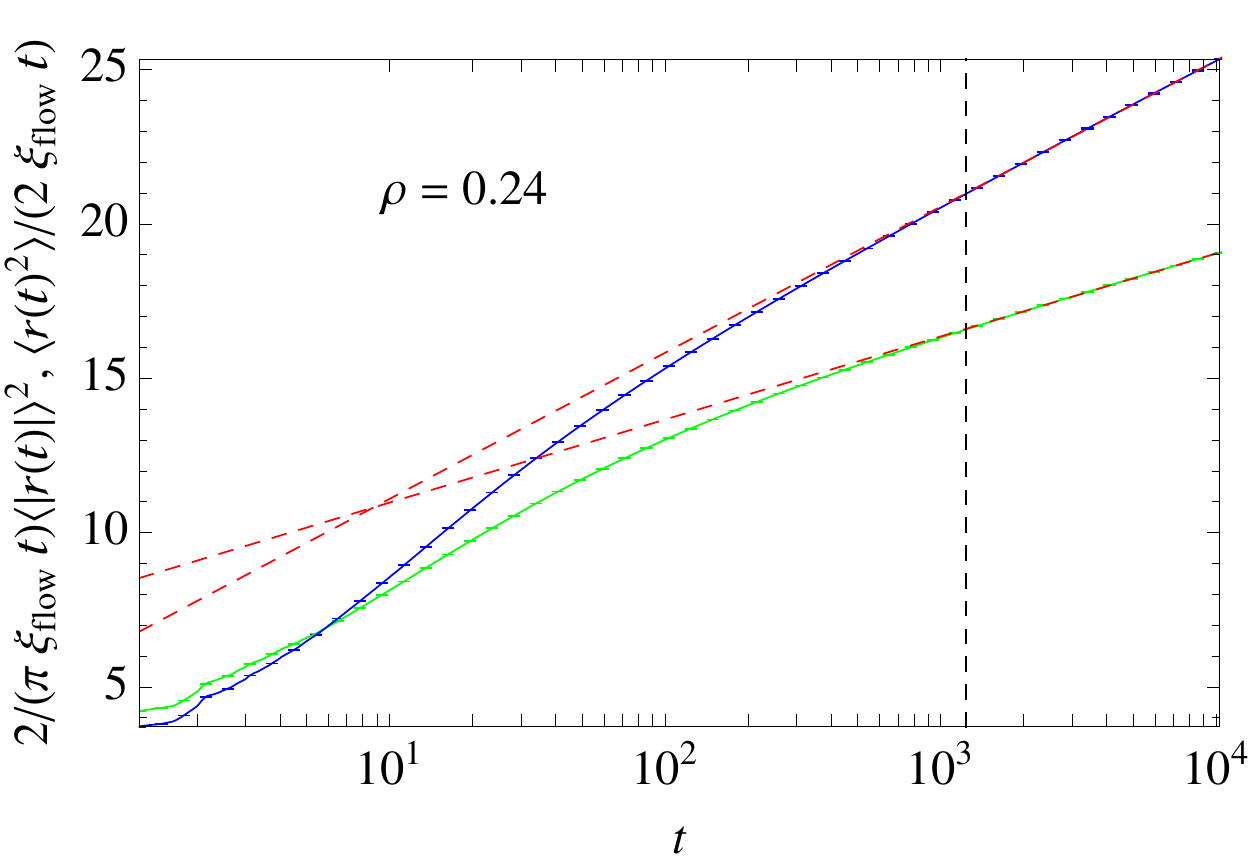}
    \label{fig.moments024}
  }
  \hfill
  \subfigure[~Intercepts, $\rho = 0.24$]{
    \includegraphics[width=.31\textwidth]{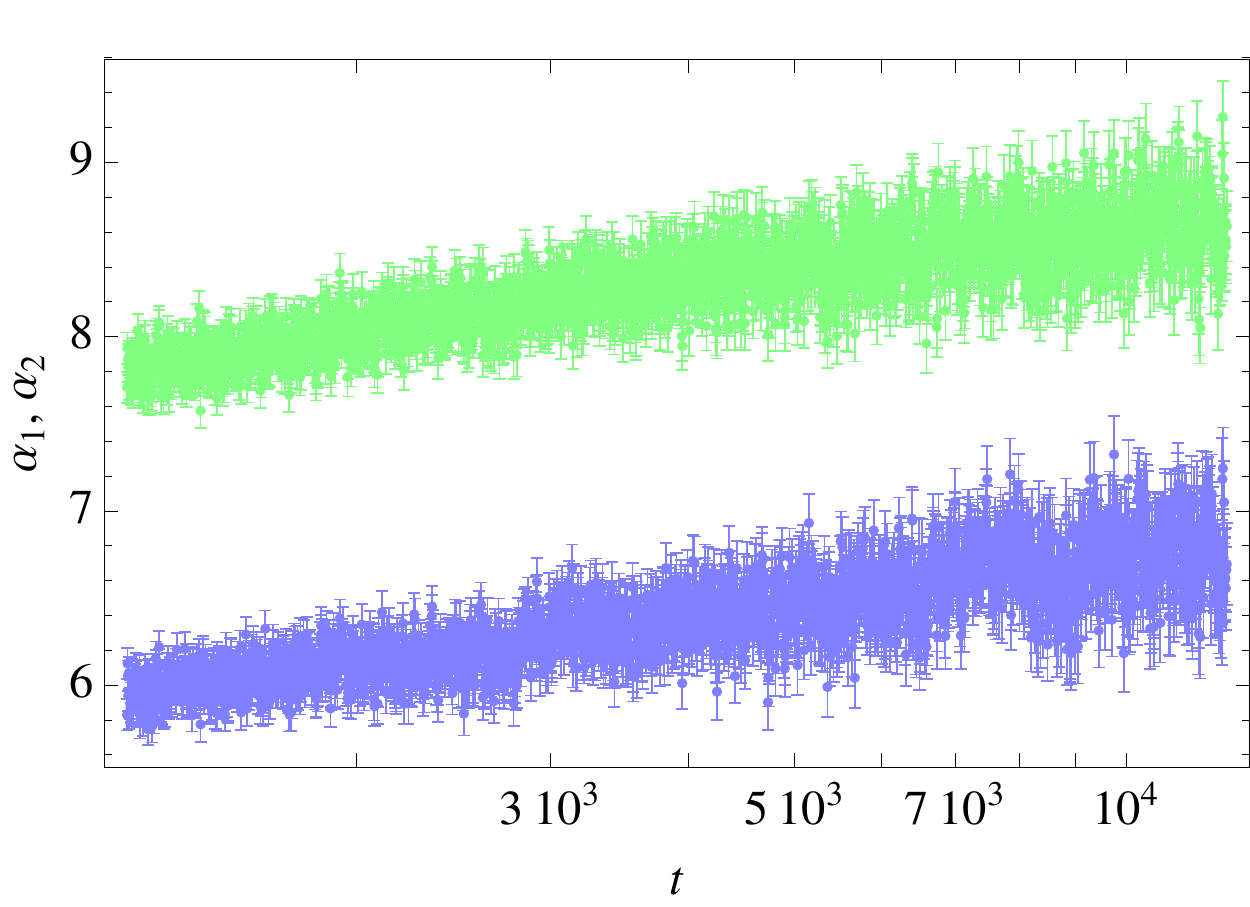}
    \label{fig.alpha024}
  }
  \hfill
  \subfigure[~Slopes, $\rho = 0.24$]{
    \includegraphics[width=.31\textwidth]{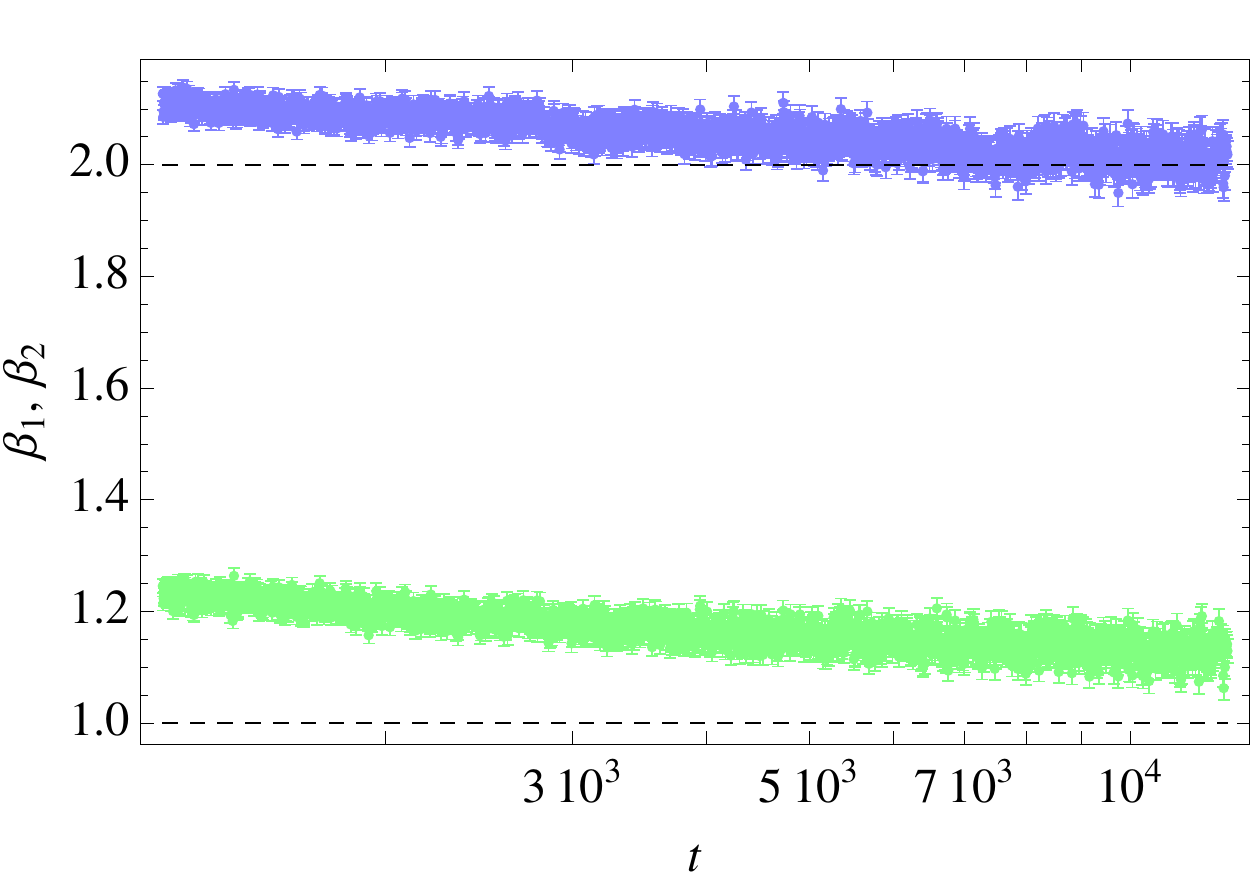}
    \label{fig.beta024}
  }
  \hfill\null
   
  \null\hfill
  \subfigure[~1st and 2nd moments, $\rho = 0.36$]{
    \includegraphics[width=.31\textwidth]{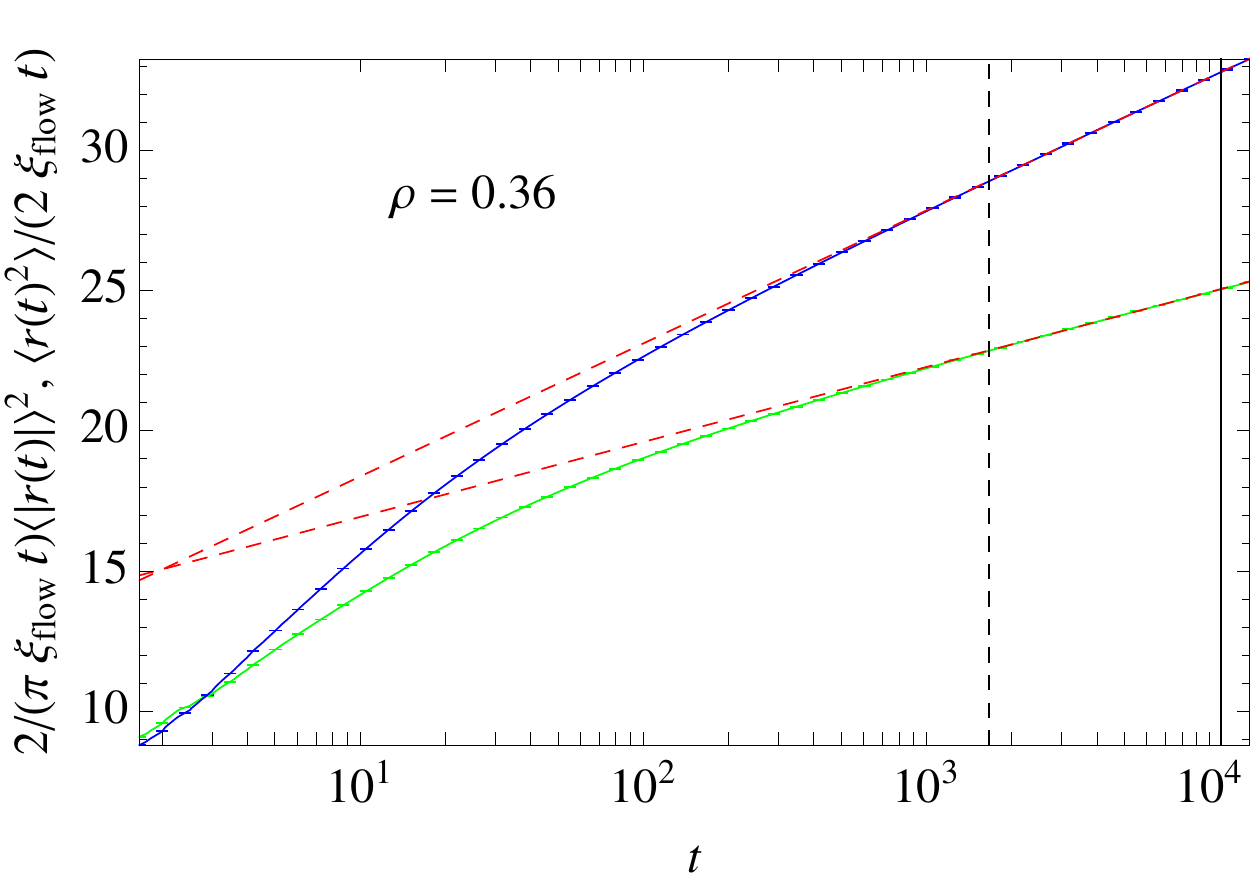}
    \label{fig.moments036}
  }
  \hfill
  \subfigure[~Intercepts, $\rho = 0.36$]{
    \includegraphics[width=.31\textwidth]{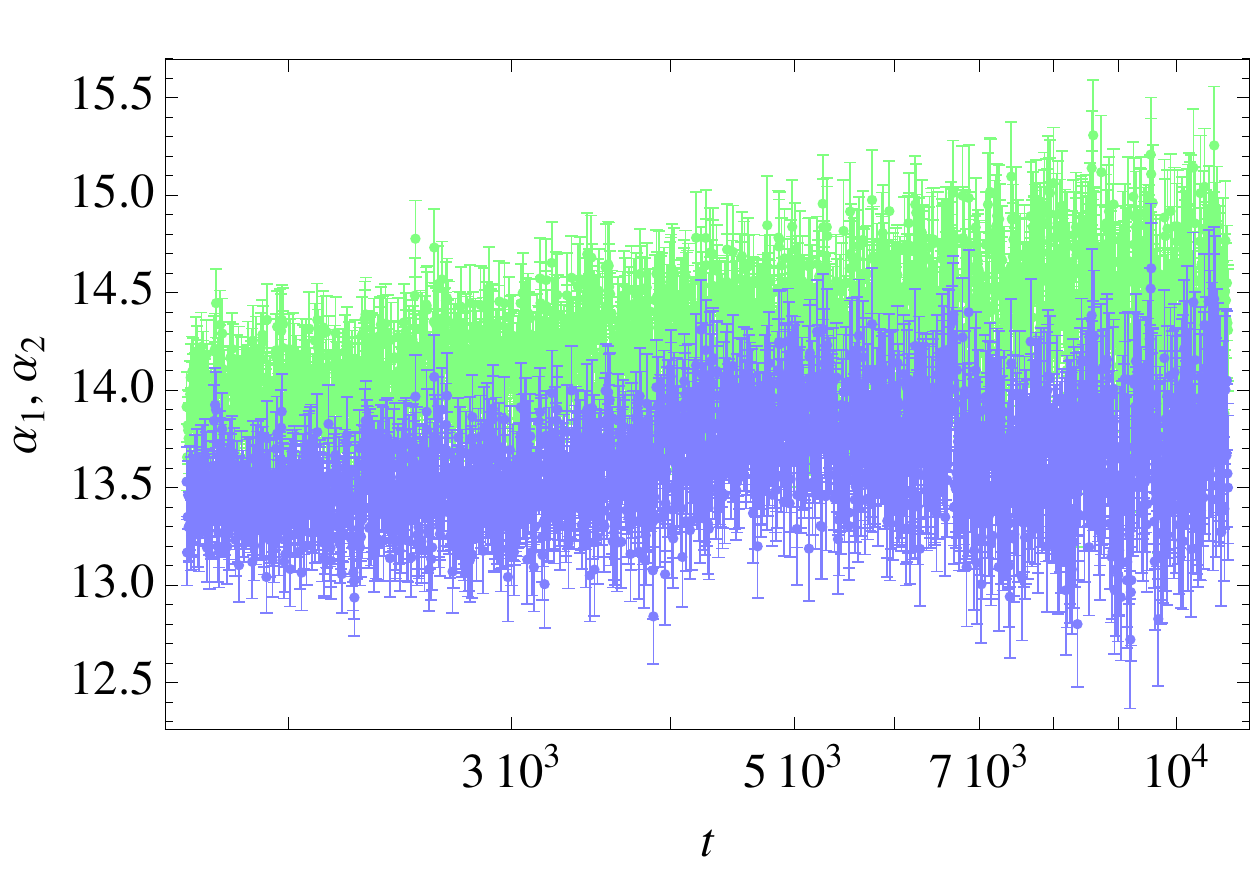}
    \label{fig.alpha036}
  }
  \hfill
  \subfigure[~Slopes, $\rho = 0.36$]{
    \includegraphics[width=.31\textwidth]{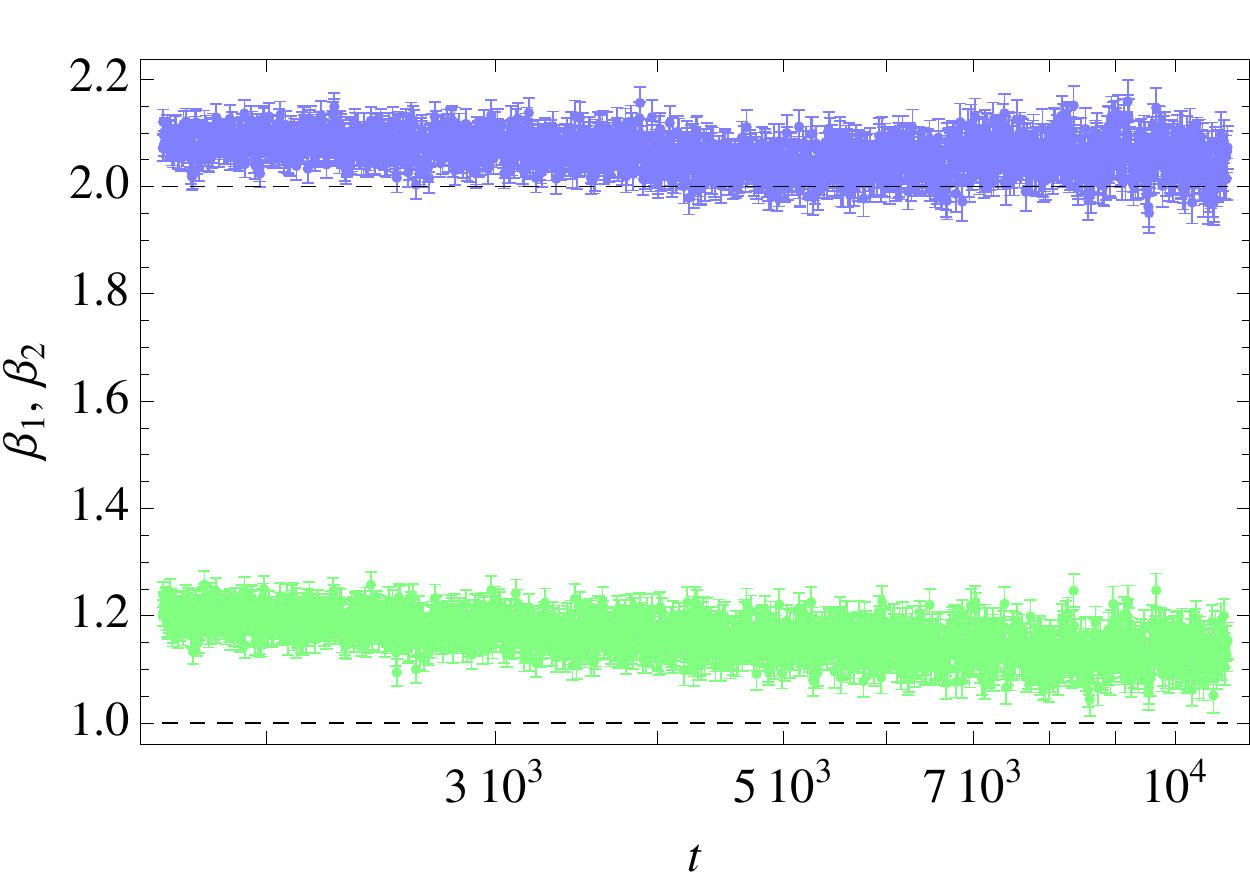}
    \label{fig.beta036}
  }
  \hfill\null
   
  \null\hfill
  \subfigure[~1st and 2nd moments, $\rho = 0.46$]{
    \includegraphics[width=.31\textwidth]{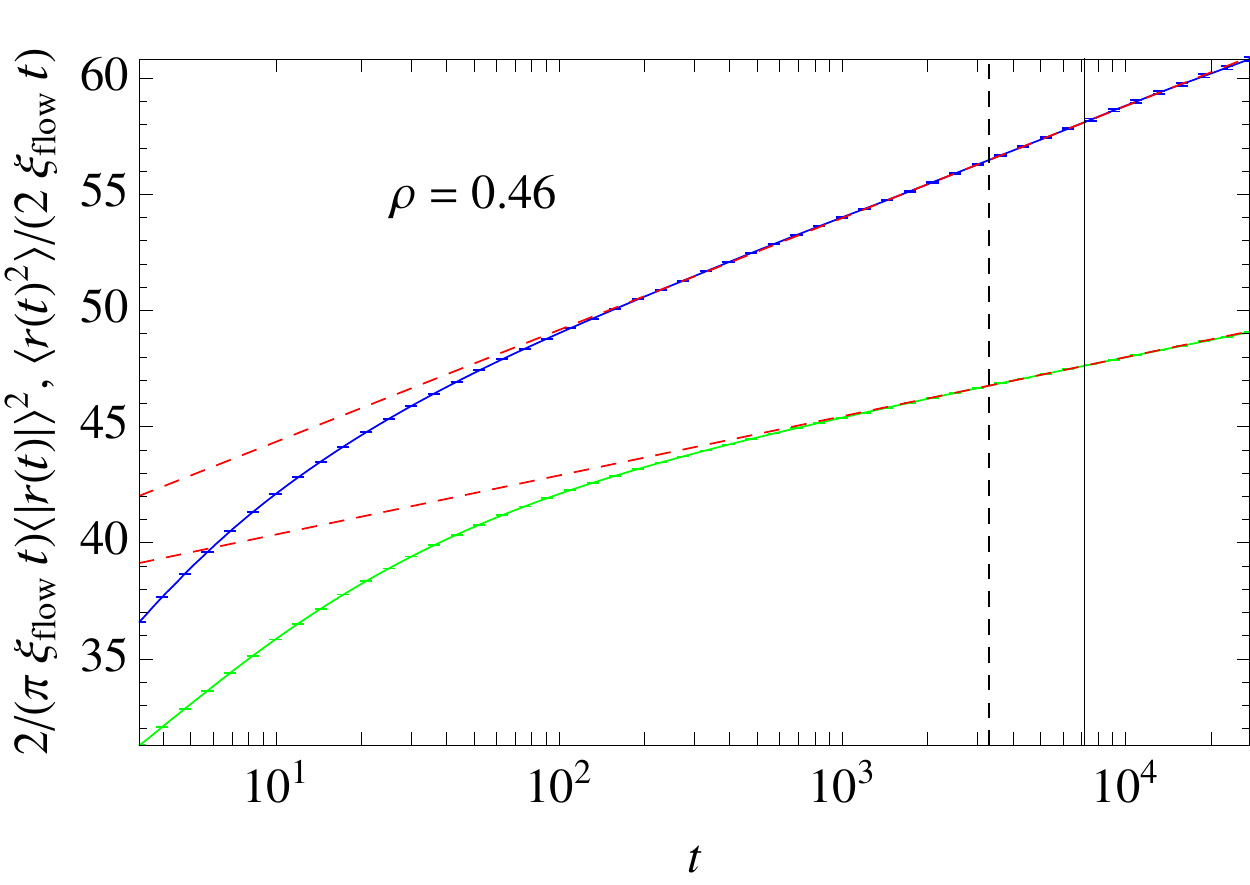}
    \label{fig.moments046}
  }
  \hfill
  \subfigure[~Intercepts, $\rho = 0.46$]{
    \includegraphics[width=.31\textwidth]{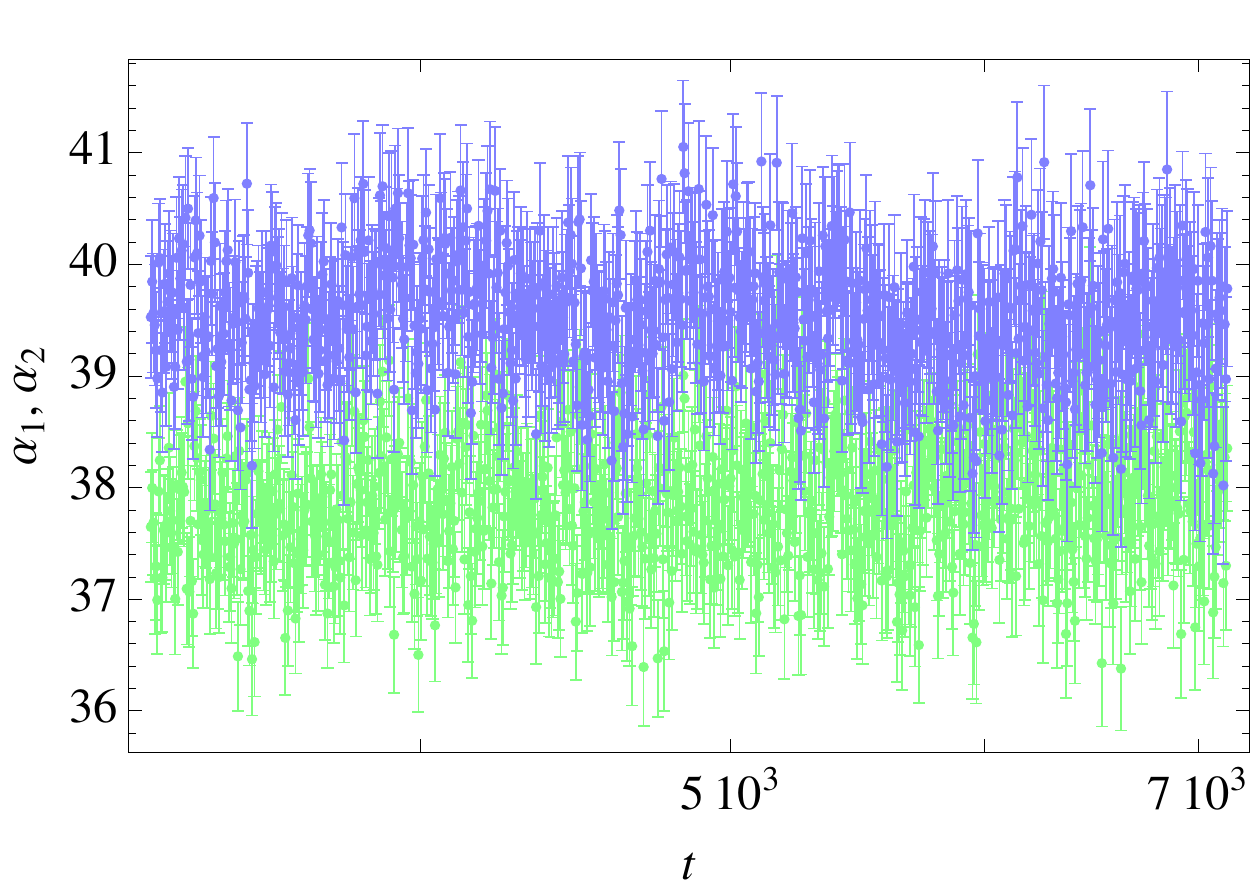}
    \label{fig.alpha046}
  }
  \hfill
  \subfigure[~Slopes, $\rho = 0.46$]{
    \includegraphics[width=.31\textwidth]{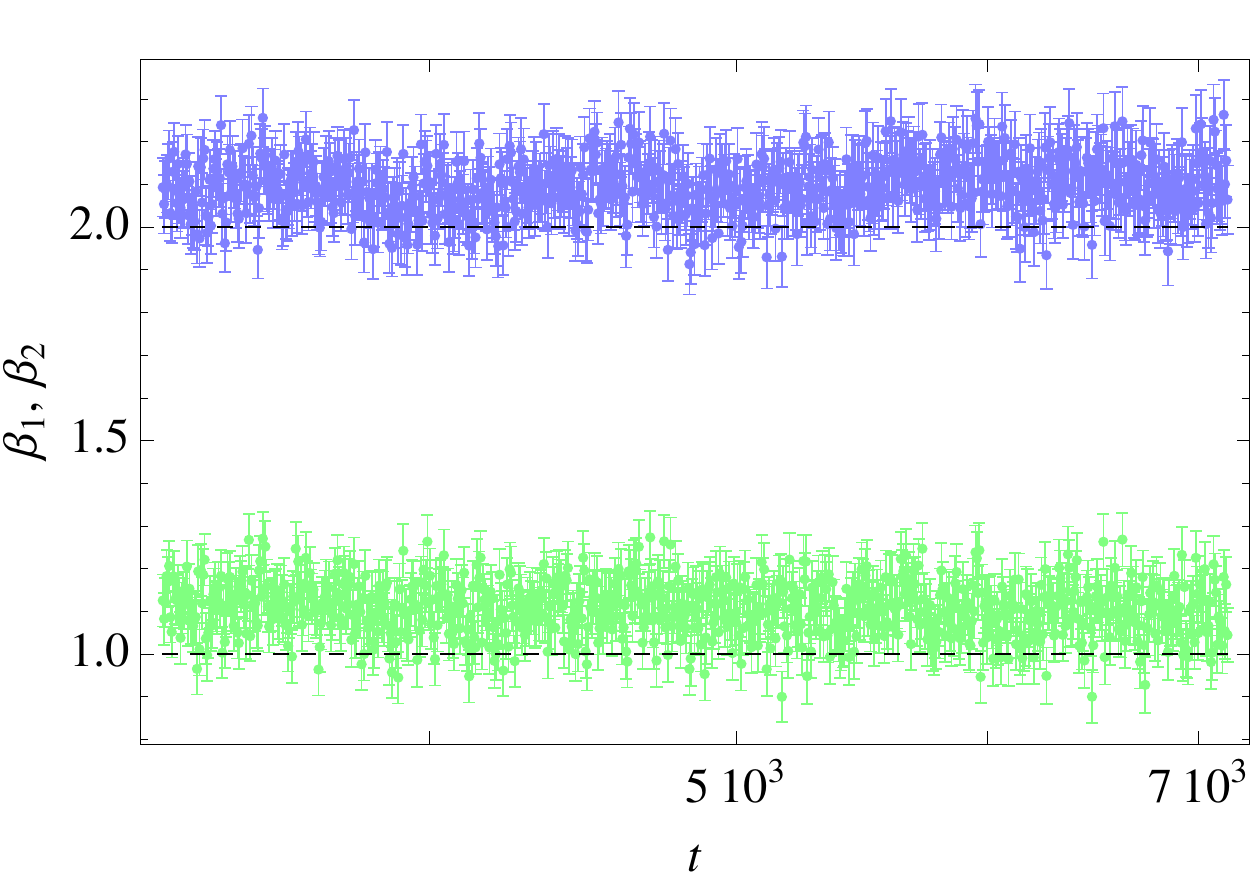}
    \label{fig.beta046}
  }
  \hfill\null
  \caption{(Color online)
    Left panels: numerical measurements of the normalized first (green, bottom
    curves) and second (blue, top curves) moments, Eqs.
    \eqref{eq:ftimemoment2}-\eqref{eq:ftimemoment1}, as functions of
    time. The dashed red lines show the results of linear fits of these
    two curves, i.e., as affine functions of $\ln t$, computed in the
    interval delimited by the two vertical lines; see
    Sec.~\ref{sec.details} for further details.
    Middle and right panels: graphs of the measured intercepts, $\alpha_i$, and
    slopes, $\beta_i$, $i=1,2$, of the normalized moments as
    functions of time, i.e., obtained by fitting straight lines
    through successive pairs of data points. The time-ranges of those
    graphs correspond to the fitting intervals shown on the left
    panels. The number of open corridor types is four for the
    parameter value $\rho = 0.14$,
    Figs.~\ref{fig.moments014}-\ref{fig.beta014}, two for $\rho = 
    0.24$, Figs.~\ref{fig.moments024}-\ref{fig.beta024}, and one 
    for $\rho = 0.36$ and $0.46$,
    Figs.~\ref{fig.moments036}-\ref{fig.beta046}. 
  }
  \label{fig.moments}
\end{figure*}

We focus on the first two
moments of the rescaled displacement vector. We are particularly
interested in the second moment for its physical relevance, but also
because the exponent $q=2$ is the onset of the anomalous behavior,
Eq.~\eqref{eq:armstead-moments}. For further validation of our
results, we provide a comparison between the second and the first
moments. As discussed earlier, we need to critically assess the effect
of \emph{finite-time} measurements of these quantities; the important
point to notice is that the ``large'' times that are needed  to
observe the asymptotics of
Eqs.~\eqref{eq:asympmoment}-\eqref{eq:moment2} must be
understood as  \emph{logarithmically} large times, i.e., times so
large that their logarithm is actually large. 

Higher moments, $q>2$, will not be analyzed here. For these moments,
it is believed that logarithmic corrections to the scaling
\eqref{eq:armstead-moments} are in fact absent
\cite{Melbourne:private}. Their measurement is, however, delicate 
\cite{Zaslavsky:1997p706, Courbage:2008p454}; we will return to
this issue in a separate publication.

Following the discussion in Sec.~\ref{sec.asymp}, we measure
numerically the left-hand side of Eq.~\eqref{eq:moment2}, which, up
to a numerical factor, is proportional to the finite-time diffusion
coefficient $D(t)$, Eq.~\eqref{eq:ftimediffcoeff}. However, as
detailed in the 
introduction, to obtain an accurate measurement of the logarithmic
divergence of this quantity, it is necessary to include terms of order
$1$ in this expression. (We do not include terms of order between
$1$ and $\log t$, as they would be invisible to our simulations.)  
Dividing by the variance so as to eliminate the
dependence on the model's parameter from the asymptotic result
\eqref{eq:moment2}, we thus seek an asymptotically affine function of
$\ln t$:  
\begin{equation}
  \frac1 {2 \xiflow t} \left \langle r(t)^{2} \right \rangle \sim
  \alpha_2+ \beta_2 \ln t,
  \label{eq:ftimemoment2} 
\end{equation}
where $\alpha_2$ and $\beta_2$ are implicit functions of time, and are
expected to converge to constant values \footnote{We are thereby
  discarding the possibility that other time-dependent terms diverging
  slower than the logarithm, such as, for instance, $\ln \ln t$, may
  be present; we do not find numerical evidence to support the
  existence of such terms.} as $t\to\infty$; according to
Eq.~\eqref{eq:moment2}, we should  find  $\beta_{2} =  2$ for large
enough times. 

In contrast to the second moment, the first moment (of the norm of the
displacement vector) must follow Eq.~\eqref{eq:asympmoment}. Taking
the square of this quantity and dividing by the variance, we again
expect an asymptotically affine function of $\ln t$: 
\begin{equation}
  \frac2 {\pi  \xiflow t} \big \langle r(t)  \big \rangle^2 \sim
  \alpha_1+ \beta_1 \ln t,
  \label{eq:ftimemoment1} 
\end{equation}
now with $\beta_{1} = 1$ for large enough times.

We refer to the quantities on the left-hand side of
Eqs.~\eqref{eq:ftimemoment2}-\eqref{eq:ftimemoment1} as the 
\emph{normalized second and first moments} respectively. 

Examples of numerical computations of these quantities are provided in
\fref{fig.moments}, for different parameter values. The left
panels display the graphs of these two normalized moments as functions
of time, on logarithmic scales. For times large enough, the curves tend
to straight lines whose fits provide estimates of the coefficients
$\alpha_i$ and $\beta_i$, shown as functions of time on the middle and
right panels, respectively; see Sec.~\ref{sec.details} for further
details on the computation of these coefficients.

As evidenced by the data shown in \fref{fig.moments}, when the linear
interpolation is performed on a (logarithmically) small neighborhood
of some given finite time $t$, we must think of all four fitting
parameters as functions of $t$.  Since the asymptotic convergences in
Eqs.~\eqref{eq:asympmoment}-\eqref{eq:moment2}  occur 
over a logarithmic time scale, it is reasonable to expect a
slow convergence of these quantities; the deviations of the
measured coefficients $\alpha_i$ and $\beta_i$ from their asymptotic
values are indeed found to decay as power laws with exponents less
than one; see the discussion below.

Let us remark once again that there are in general no analytical
predictions for the fitting parameters $\alpha_i$. Though the slopes
$\beta_i$ ($i=1,2$) are asymptotically independent of the model's
parameter $\rho$,  this is not expected of the intercepts $\alpha_i$;
using dimensional arguments, it is in fact not difficult to convince
oneself that $\alpha_i$ should diverge with $1/(1 - 2 \rho)$ as $\rho
\to 1/2$. In other words, $\alpha_i \gg \beta_i$ in the limit of
narrow corridors.  At times $t$ attainable in numerical simulations
we should thus typically expect the intercepts to be of sizes similar
to the terms $\beta_i \ln t$ on the right-hand sides of
Eqs.~\eqref{eq:ftimemoment2}-\eqref{eq:ftimemoment1}, or even much
larger, as occurs when $\rho\to1/2$; this regime will be analyzed in a
separate publication \cite{us:short}.

We believe this observation is key to explaining the difficulties met
in observing numerically the asymptotic scalings in  
Eqs.~\eqref{eq:asympmoment}-\eqref{eq:moment2}.   
Recognizing that short time averages tend to be dominated by
diffusive motion helps explain the relevance of the fitting parameters
$\alpha_i$. Indeed the effect of ballistic trajectories on the
statistics of displacements is feeble and the anomalous logarithmic
divergences have a rather weak influence on the finite-time statistics,
especially so when $\rho$ is close to its upper bound, $1/2$ (so that
the horizontal and vertical corridors have narrowing widths).

\subsection{Time span of measurements}

For a given value of the parameter $\rho$, a key issue is to
determine a time interval where the fitting parameters
$\alpha_i$ and $\beta_i$ in
Eqs.~\eqref{eq:ftimemoment2}-\eqref{eq:ftimemoment1} can be accurately
measured.  As discussed in the introduction, integration times should
be neither too short, nor too 
long; they should be large enough to avoid the regime where transient
effects dominate, but cannot be too large, since the number of initial
conditions required to sample the moments up to a given time scale
grows with the square of this scale; see the discussion below.

At the level of the dynamics, there are distinct time scales at
play. The first is the mean free time, $\taumft $,
Eq. \eqref{eq:mftime}, which measures the average time that
separates successive collisions with scatterers. 

To identify a second timescale, which characterizes the motion of a
point particle on the billiard table, consider a lattice of unit cells
each shifted by one half of a unit length in both vertical and
horizontal directions, so that obstacles are now sitting at the cells'
corners rather than at their centers.  The average time it takes for a
particle at unit speed to exit a cell after it entered it (with
position and velocity distributed according to the Liouville measure
for the Poincar\'e section given by the four line segments delimiting
each cell) is \cite{Chernov:1997p1}
\begin{equation}
  \label{eq:tauex}
  \tauex = \frac{\pi ( 1 - \pi \rho^2)}{ 4 (1 - 2 \rho)}.
\end{equation}
The residence time $\tauex$ provides a natural time unit of lattice
displacements. Note that, whereas $\taumft$, Eq.~\eqref{eq:mftime},
diverges in the limit of small scatterers, $\rho \to 0$, $\tauex$
diverges in the opposite limit of narrow corridors, $\rho \to 1/2$. 

In the presence of infinite corridors, the other relevant timescale is
of course the ballistic one, which, for horizontal and vertical
corridors, is, in the units of cell sizes and speed of point particles,  
\begin{equation}
  \tauball = 1.
  \label{eq:tauball}
\end{equation}

\begin{figure*}[htb]
  \centering
  \null\hfill
  \subfigure[~Distribution of ballistic segments, $\rho = 0.14$]{
    \includegraphics[width=.45\textwidth]{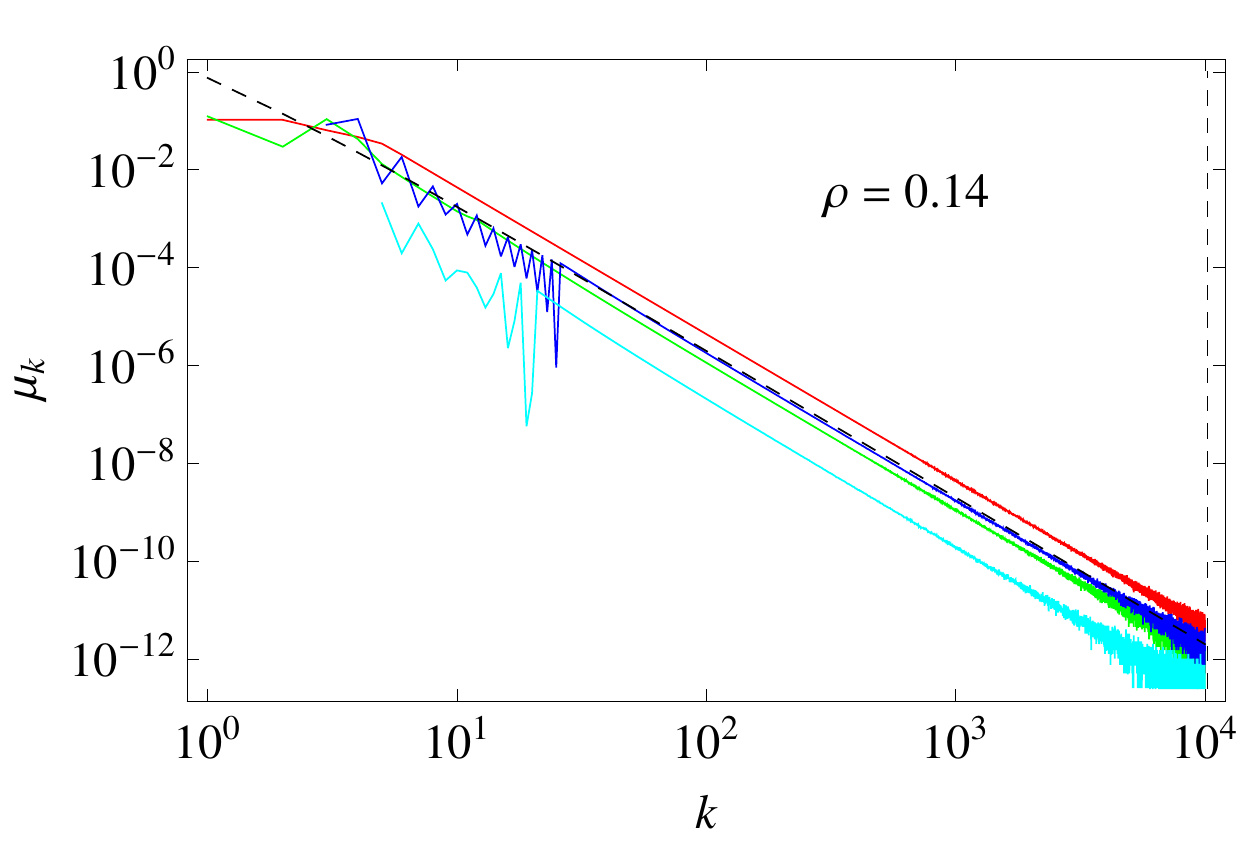}
    \label{fig.mu014}
  }
  \hfill
  \subfigure[~Cumulative distribution of ballistic segments, $\rho = 0.14$]{
    \includegraphics[width=.45\textwidth]{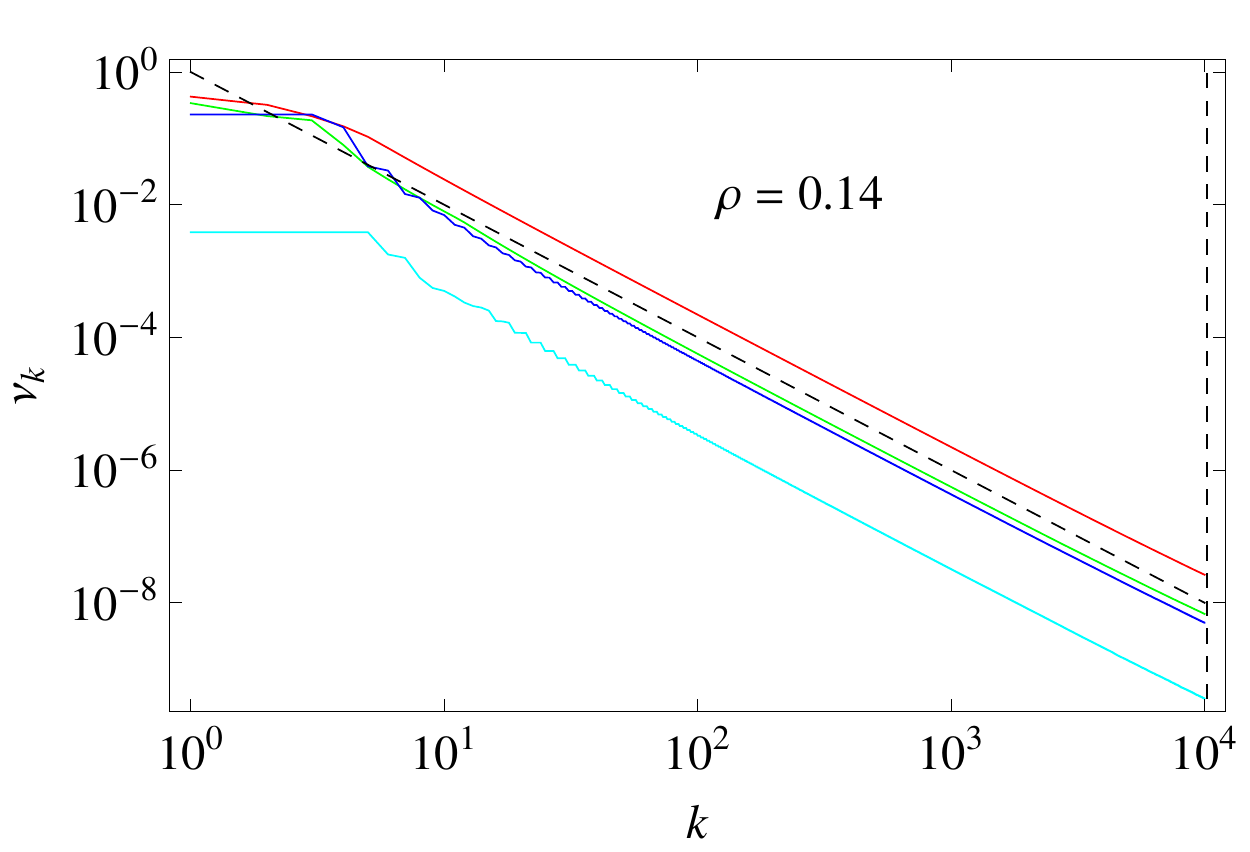}
  \label{fig.nu014}
  }
  \hfill\null
  \caption{(Color online)
    Numerical measurements of the distributions of the ballistic
    segments for the parameter value $\rho = 0.14$. Four different
    types of corridors are open: $(1,0)$ (red curve), $(1,1)$ (green
    curve), $(1,2)$ (blue curve), and $(1,3)$ (cyan curve). The black
    dashed curves are defined according to
    Eqs.~\eqref{eq:muk}-\eqref{eq:nuk}, respectively, and have the
    same asymptotics. 
  }
  \label{fig.munu}
\end{figure*}

As can be seen in \fref{fig.moments}, the initial regime, which is dominated
by  transient effects, is typically longer for the first moment than the 
second. For either moment, however, the lengths of
transients may depend on several 
factors, such as the persistence of  correlations, or the precise
distribution of ballistic segments, which vary with the parameter's
value. Moreover, in an idealized model such that ballistic segments
are independent and identically distributed (so that correlations are
absent) $\alpha_i$ and $\beta_i$ display $1/t$  corrections
to their asymptotic values \cite{us:long}. For the billiard dynamics,
we should a fortiori expect a decay not faster than $1/t$; the convergence
to the asymptotic regime might be even slower.

The distinction between the initial transient regime and that where
the moments follow
Eqs.~\eqref{eq:ftimemoment2}-\eqref{eq:ftimemoment1}, with
fitting parameters displaying small finite-time corrections, is
not sharp. There is in fact no easy way to estimate the length of the
initial transient regime, a priori. Empirically, however, we find the
initial regime to subdue after a duration between $10^3$ and $10^4$
units of  $\taumft$, which, in the range of parameter values $\rho$
investigated, may thus vary from several hundred units of $\tauex$ to
a few thousand, depending on the value of the parameter.   

Integration times should therefore be greater than a few thousand units
of $\tauex$. How much greater depends on one's ability to
sample ballistic segments over that duration. 
Since it is known \cite{Szasz:2007v129p59} that the 
distribution of the lengths of ballistic segments
in infinite-horizon periodic Lorentz gases decays with the cube of
their lengths, 
\begin{equation}
  \mu_k 
  \sim \frac1{k^3},
  \label{eq:muk}
\end{equation}
let us compare our numerical findings to the following 
model for the corresponding cumulative distribution, by which we mean
the probability of having ballistic segments of lengths at least $k$, 
\begin{equation}
  \nu_k = \sum_{j = k}^\infty \mu_j
  = \frac{1}{k^2}.
  \label{eq:nuk}
\end{equation}
Incidentally, $\nu_1 = 1$ is equivalent to normalization of the
probabilities \eqref{eq:muk}, for $k$ ranging over positive integers.
The $k^2$ decay in the above equation is
responsible for the logarithmic divergence of
the normalized  second moment.

Now, suppose that, for times up to $k$, we want to accurately
sample ballistic segments of lengths $k$---assuming that this
will imply a good sampling of shorter ballistic segments as well.
Since  point particles have unit speed, the flights of lengths $k$
cannot experience a collision in the time interval $[0,k]$ (modulo
corrections that are  
negligible for large $k$). They are thus indistinguishable, in the
time-frame considered, from flights of lengths $\ge k$, as we
do not know when the last collision before time 0 occurred 
or when the first one after time $k$ will. So we must consider 
the entire set of ballistic segments of lengths $\ge k$, whose
probability is given by \eqref{eq:nuk}, as
a whole, and each segment there will contribute with a length
$k$, in the time interval $[0,k]$.
Accordingly, the number of initial
conditions we have to sample in order to accurately measure the
moments for times up to $k$ grows like $k^2$.   

The tails of the actual distributions of ballistic segments, shown in
\fref{fig.munu} for $\rho = 0.14$, differ from
Eqs.~\eqref{eq:muk}-\eqref{eq:nuk} by a numerical factor which is
$\mathcal{O}(1-2\rho)$. In the narrow corridor limit, transitions
between cells are overwhelmingly dominated by segments of unit
lengths; ballistic segments are thus infrequent. Away from this limit,
however, ballistic trajectories occur at manageable frequencies.  

Considering the examples displayed in \fref{fig.moments}, the
initial transient regime subdues after times roughly $t\sim 10^4$. For
the parameter values shown, and up to a constant factor of order
unity \footnote{This statement applies to type $(0,1)$ corridors. The
  frequency of occurrence of corridors of other types changes
  according to their relative widths.}, the cumulative distribution of
ballistic segments is well approximated by Eq.~\eqref{eq:nuk}, at
least for $k\gg1$. We thus typically need $10^{8}$ initial conditions
to sample trajectories up to time $t \sim 10^{4}$, but would need
$10^{10}$ initial conditions to sample trajectories for times up to $t
\sim 10^{5}$. While the former is within reach of our numerical
computations, the latter is not, at least not unless one is able to
devote about ten years' worth of CPU time to it, notwithstanding the
impact of the limited accuracy of the integration on the results. 

Statistical averages are, in practice, limited to ensembles
of at most a billion trajectories of duration $10^4 \tauex$. The
margin between the decay of initial transients and the largest timescale for
which ballistic trajectories are accurately sampled is thus typically
narrow. 

\subsection{Details of the numerical procedure and results}
\label{sec.details}

\begin{table*}[hbt]
  \begin{tabular}{|c||c|c|c||c|c||c|c|}
    \hline
    $\rho$ & $t_\mathrm{tot}$ & $t_\mathrm{f}$ & $N$ &
    $\alpha_1$ &  $\beta_1$ &  
    $\alpha_2$ &  $\beta_2$ \\
    \hline
    $0.14$ & 
    $1.02\times 10^{5}$ &
    $1.65\times 10^{4}$ &
    $240$ &
    $6.8 \pm 0.6$ &
    $1.11 \pm 0.06$ &
    $3.8 \pm 0.5$ &
    $2.02 \pm 0.05$ 
    \\
    & 
    $1.02\times 10^{4}$ &
    $1.02\times 10^{4}$ &
    $1440$ &
    $6.50 \pm 0.02$ &
    $1.142 \pm 0.002$ &
    $3.67 \pm 0.07$ &
    $2.029 \pm 0.007$ 
    \\
    \hline
    $0.24$ & 
    $1.24\times 10^{4}$ &
    $1.24\times 10^{4}$ &
    $1000$ &
    $8.58 \pm 0.08$ &
    $1.135 \pm 0.009$ &
    $6.68 \pm 0.02$ &
    $2.0059 \pm 0.0007$ 
    \\
    \hline 
    $0.36$ & 
    $1.66\times 10^{4}$ &
    $1.10\times 10^{4}$ &
    $1000$ &
    $14.33 \pm 0.03$ &
    $1.152 \pm 0.003$ &
    $13.54 \pm 0.05$ &
    $2.069 \pm 0.005$ 
    \\
    \hline
    $0.46$ & 
    $3.29\times 10^{4}$ &
    $7.16\times 10^{3}$ &
    $1200$ &
    $37.8 \pm 0.7$ &
    $1.10 \pm 0.08$ &
    $39.8 \pm 0.6$ &
    $2.11 \pm 0.06$ 
    \\
    & 
    $3.29\times 10^{3}$ &
    $3.29\times 10^{3}$ & 
    $1100$ &
    $37.44 \pm 0.06$ &
    $1.150 \pm 0.008$ &
    $40.24 \pm 0.05$ &
    $1.994 \pm 0.007$ 
    \\
    \hline
  \end{tabular}
  \caption{Numerical measurements of the fitting parameters $\alpha_i$
    and $\beta_i$, $i=1,2$, of the normalized moments
    \eqref{eq:ftimemoment2}-\eqref{eq:ftimemoment1} for different values
    of the model's parameter, $\rho$. For each parameter value, we
    indicate $t_\mathrm{tot}$, the total integration time, 
    $t_\mathrm{f}$ the largest time in the fitting range, and $N$, the number of groups of $10^6$
    trajectories used to gather the corresponding data. The values of
    the coefficients $\alpha_i$ and $\beta_i$, $i= 1,2$ reported here
    are obtained by computing the means and their standard deviations
    for measurement times $t$ near $t_\mathrm{f}$.
  }
  \label{tab.results}
\end{table*}

Numerical integration of trajectories on the infinite-horizon
Lorentz gas proceeds according to standard event-driven algorithms,
which are common to systems with hard-core interactions
\cite{Lubachevsky:y1991v94p255}. A specificity, however, is in the
choice of initial conditions, which are sampled from the standard
Liouville distribution along the vertical and horizontal borders of a unit
cell. This choice allows to sample trajectories initially in the
process of completing a ballistic segment, as part of the equilibrium
distribution; such trajectories would be absent from the statistical
ensemble had we instead chosen to distribute initial conditions on the
surface of the obstacle, i.e., where collision events take place. 

Each trajectory is integrated over a given time span $t_\mathrm{tot}$, during which the
particle's position on the billiard table is regularly sampled, at
intervals of time uniformly distributed on a logarithmic 
scale. Statistical averages of observables such as the normalized moments
\eqref{eq:ftimemoment2}-\eqref{eq:ftimemoment1} are obtained by
repeating measurements of these quantities for a large number of
initial conditions. 

As integration proceeds, the distribution of ballistic segments is
computed by recording successive ballistic segments, according to
their lengths and corridor types, see \fref{fig.mu014}. At the end of
the integration, we obtain a criterion for determining an upper bound 
$t_\mathrm{f}$, 
of the times up to which statistical averages are reliably computed:
we  choose $k_{\mathrm{f}}$ to be the smallest integer $k$ such that
the percentage of unsampled  ballistic segments of lengths $\le k$ 
exceeds $0.1\%$.
Finally we set $t_{\mathrm{f}}=k_{\mathrm{f}}$, as speed is fixed to one. 
On the left panels of \fref{fig.moments}, $t_\mathrm{f}$ is marked by
vertical solid lines. Of course, this is but a rough way of estimating the
largest ballistic scale such that the cumulative distribution of
ballistic segments is accurately sampled. Though it may not be
optimal,  this is a quantitative  criterion that has, among its
advantages, that it is easy to implement and requires no a priori
knowledge of the distribution of ballistic segments.

In \fref{fig.moments}, the lower bounds of the fitting intervals,
marked by solid vertical lines on the left panels, are taken to be
one tenth of the total integration time of the
simulation, $t_\mathrm{i} \equiv t_\mathrm{tot}/10$, which we set
to $t_\mathrm{tot} \equiv 10^4 \tauex$. The bounds 
$t_\mathrm{i}$ and $t_\mathrm{f}$ also correspond to
the time ranges of the figures shown on the middle and right
panels of \fref{fig.moments}. The data themselves are 
obtained by computing the averages of the normalized first and second
moments \eqref{eq:ftimemoment2}-\eqref{eq:ftimemoment1} as functions
of time for about one thousand groups of $10^6$ trajectories each; see
Table \ref{tab.results} for details. The
times $\{t_k\}_{k=1}^{N}$, at which the moments are computed, span
$N=10^4$ sub-intervals of $t_\mathrm{tot}$, such that $t_{k}/t_{k-1} =
N^{1/N}$ and $t_N = t_\mathrm{tot}$. 

Having determined fitting intervals by inspection of the distributions
of ballistic segments according to the criterion described above, we
compute, for each measurement time $t_k$ within the fitting interval,
$t_\mathrm{i} \leq t_k \leq t_\mathrm{f}$, the values of the fitting parameters
$\alpha_i$ and  $\beta_i$ of the normalized moments averaged over a
group of $10^6$ trajectories. This is achieved by fitting straight lines through
successive pairs of data points, at $t_k$ and  $t_{k+1}$. Since the
measurement times are spread uniformly on a logarithmic scale, we
obtain in this way, for the set of measurement times $t_k$ in the
interval between $t_\mathrm{i}$ and $t_\mathrm{f}$, sequences of
values of $\alpha_i$ and $\beta_i$, whose means (for
the corresponding measurement times $t_k$) are displayed on the middle
and right panels of \fref{fig.moments}. The standard deviations of
these means yield the   
corresponding error bars.  

In Table \ref{tab.results}, we extracted from \fref{fig.moments} the
values of the fitting parameters $\alpha_i$ and $\beta_i$, measured at
the right-ends of the fitting intervals. The precision reported on
those values reflect the fluctuations observed over the last ten data
points of each of the fitting intervals. Increasing the time span
would clearly result in smaller error bars. It must however be assumed
that the fitting parameters do not exhibit significant time dependence
over the chosen time span. 

Integration times vary with the value of the model's parameter
$\rho$. For small values of $\rho$ it is possible 
to take longer integration times, with large enough numbers of initial
conditions. On the contrary, when $\rho$ increases towards $1/2$,
integration times have to be decreased in order to allow for large
enough numbers of initial conditions. For $\rho = 0.14$, we also 
report in Table \ref{tab.results} values of the fitting parameters
obtained by integrating over a total time of $10^5\tauex$. The width
of the fitting interval is thus larger than that obtained by integrating
over times $10^4\tauex$, but only by a small factor; accordingly, the
values of the fitting parameters do not vary appreciably.

At the opposite end of the range of parameter values shown here, for
$\rho = 0.46$, though the number of initial conditions reported for
integration times up to $10^4\tauex$ is rather large, the precision on
the fitting  parameters is not as good, particularly for $\beta_2$;
this is also reflected by the fluctuations observed in the data
displayed in \fref{fig.beta046}. Repeating the measurement over a
total integration time of $10^3\tauex$, we obtain better statistics
for comparable fitting times. 

Overall, the convergence of $\beta_2$ to its asymptotic value, $2$, is
observed with better accuracy than that of $\beta_1$ to $1$. The values
obtained are consistent throughout the range of the model's parameter
values, in spite of the variations in the values of the intercepts,
$\alpha_1$ and $\alpha_2$. We interpret this as a clear vindication of
our methods; weak logarithmic divergences of the mean-squared
displacements of point-particles on infinite-horizon billiard tables
can be measured with satisfactory precision, regardless of their
strength, gauged by the variance $\xiflow$,
Eq.~\eqref{eq:variancetotal}.

\section{Conclusions
  \label{sec:conclusion}
}

The periodic Lorentz gases on a square lattice investigated in this
paper are prototypical examples of infinite-horizon billiard tables,
exhibiting a weak from of super-diffusion. Such a regime is marginal
in the sense that it lies at the border between  regimes of normal
diffusion and regimes of anomalous super-diffusion with mean-squared
displacement  growing with a power of time strictly greater than unity. 

In the case of our ``weak super-diffusion", corrections to the linear growth
of the mean-squared displacement are logarithmic in time. For
moderately large times, i.e., those times which are accessible to
numerical computations, the slow growth of these corrections implies 
the coexistence of  two distinct regimes, one of normal diffusion,
whereby point-particles move short distances between collisions with
obstacles, i.e., of order of the inter-cell distance,  and, the other,
a regime of accelerated (also termed enhanced) diffusion due to the
presence of ballistic trajectories.  

Though the asymptotic regime---that which exhibits the logarithmic
divergence---has been well understood on a rigorous level, much less
can be said about the regime of normal diffusion with which it
typically coexists. As argued in this paper, ignoring this second
regime ultimately masks the asymptotic regime itself, precluding its
accurate detection. In this respect, a great deal can be learned from a
careful numerical investigation.

The analysis presented in this paper has focused on two moments of the
normalized displacement, each with distinct characteristics. On
one hand, the first moment (of the modulus) of the displacement vector
was taken as a benchmark of the limit law
\eqref{eq:asympmoment}, for which we could check the convergence
of the corresponding moment of the anomalously rescaled
process. Though this convergence could be verified 
with good accuracy throughout the range of the model's parameter
values we investigated, it must be noted that it appears to be slower
in the regime of large corridors than in the opposite regime, of
narrow corridors.  

The second moment, on the other hand, is of particular importance
because it marks the onset of the anomalous scaling regime
\eqref{eq:armstead-moments}. As noted, the corresponding moment of the
anomalously rescaled process is expected to converge to twice its
limiting variance. This observation is indeed consistent with
numerical measurements of this quantity, to within very good accuracy 
in most cases. 

The numerical investigations reported in this paper are based on
standard event-driven algorithms, with uniform sampling of
trajectories. No attempt was made to use special techniques to improve
the sampling of ballistic trajectories. Further 
investigations will focus on refined algorithms specifically designed
to explore phase-space regions associated with such rare events, e.g., in the spirit
of Refs.~\cite{hsu:2011review, laffargue:2014locating,
  leitao:2013monte}, and assess their usefulness for the sake of 
computing statistical averages.

A separate perspective relates to the connection between infinite
Lorentz gases and stochastic processes. The infinite-horizon Lorentz
gas can indeed be viewed as an example of a correlated L\'evy walk,
whose distribution of free paths scales with the inverse cubic power
of their lengths. Models of such walks appear in the context of random
search algorithms  \cite{viswanathan:2011physics}. Better understood,
however, are uncorrelated L\'evy walks \cite{Geisel:1985p8023,
  Zumofen:1993p804}. In this context scalings of the mean-squared
displacement such as Eq.~\eqref{eq:ftimediffcoeff} are known to occur
when the free paths are distributed as in Eq.~\eqref{eq:muk}. In a
separate publication, we will show that the narrow
corridor limit of the infinite-horizon Lorentz gas is a fertile study
ground for a class of such walks, where both normal and anomalous
diffusion coexist. Much in the spirit of the Machta-Zwanzig
approximation to the diffusion coefficient of normally diffusive 
finite-horizon periodic billiard tables \cite{Machta:1983p182},
correlations between successive ballistic segments die out as the
narrow corridors limit is reached. In this limit, the terms of order $1$ in the
normalized moments take on a simple dimensional 
form which plainly justifies the contrast between the coexistence of
normally and anomalously diffusive contributions in the finite-time
expression of the mean-squared displacement. To describe this limit in
an appropriate framework, we will introduce a description in terms of
continuous time random walks with delay. As it turns out, this is but a
particular case of a much larger class, which includes models with
all diffusive regimes, ranging from sub- to super-diffusive. The
details will be reported elsewhere.

\begin{acknowledgments}
We thank D. Sz\'asz for helpful comments, in particular with
regards to the derivation of \eqref{eq:variancetotal}, as well as  
N. Chernov and I. Melbourne for sharing unpublished results. This work
was partially supported by FIRB-project RBFR08UH60 {\em Anomalous transport of
  light in complex systems} (MIUR, Italy), by SEP-CONACYT  
grant CB-101246 and DGAPA-UNAM PAPIIT grant IN117214 (Mexico), and by
FRFC convention 2,4592.11  (Belgium). TG is  
financially supported by the (Belgian) FRS-FNRS.   
\end{acknowledgments}


\begin{thebibliography}{58}%
\makeatletter
\providecommand \@ifxundefined [1]{%
 \@ifx{#1\undefined}
}%
\providecommand \@ifnum [1]{%
 \ifnum #1\expandafter \@firstoftwo
 \else \expandafter \@secondoftwo
 \fi
}%
\providecommand \@ifx [1]{%
 \ifx #1\expandafter \@firstoftwo
 \else \expandafter \@secondoftwo
 \fi
}%
\providecommand \natexlab [1]{#1}%
\providecommand \enquote  [1]{``#1''}%
\providecommand \bibnamefont  [1]{#1}%
\providecommand \bibfnamefont [1]{#1}%
\providecommand \citenamefont [1]{#1}%
\providecommand \href@noop [0]{\@secondoftwo}%
\providecommand \href [0]{\begingroup \@sanitize@url \@href}%
\providecommand \@href[1]{\@@startlink{#1}\@@href}%
\providecommand \@@href[1]{\endgroup#1\@@endlink}%
\providecommand \@sanitize@url [0]{\catcode `\\12\catcode `\$12\catcode
  `\&12\catcode `\#12\catcode `\^12\catcode `\_12\catcode `\%12\relax}%
\providecommand \@@startlink[1]{}%
\providecommand \@@endlink[0]{}%
\providecommand \url  [0]{\begingroup\@sanitize@url \@url }%
\providecommand \@url [1]{\endgroup\@href {#1}{\urlprefix }}%
\providecommand \urlprefix  [0]{URL }%
\providecommand \Eprint [0]{\href }%
\providecommand \doibase [0]{http://dx.doi.org/}%
\providecommand \selectlanguage [0]{\@gobble}%
\providecommand \bibinfo  [0]{\@secondoftwo}%
\providecommand \bibfield  [0]{\@secondoftwo}%
\providecommand \translation [1]{[#1]}%
\providecommand \BibitemOpen [0]{}%
\providecommand \bibitemStop [0]{}%
\providecommand \bibitemNoStop [0]{.\EOS\space}%
\providecommand \EOS [0]{\spacefactor3000\relax}%
\providecommand \BibitemShut  [1]{\csname bibitem#1\endcsname}%
\let\auto@bib@innerbib\@empty
\bibitem [{\citenamefont {Chernov}\ and\ \citenamefont
  {Markarian}(2006)}]{Chernov:2006p683}%
  \BibitemOpen
  \bibfield  {author} {\bibinfo {author} {\bibfnamefont {N.}~\bibnamefont
  {Chernov}}\ and\ \bibinfo {author} {\bibfnamefont {R.}~\bibnamefont
  {Markarian}},\ }\href@noop {} {\emph {\bibinfo {title} {{Chaotic
  Billiards}}}},\ \bibinfo {series} {Mathematical Surveys and Monographs},
  Vol.\ \bibinfo {volume} {127}\ (\bibinfo  {publisher} {American Mathematical
  Society, Providence, RI},\ \bibinfo {year} {2006})\BibitemShut {NoStop}%
\bibitem [{\citenamefont {Gaspard}(1998)}]{Gaspard:1998book}%
  \BibitemOpen
  \bibfield  {author} {\bibinfo {author} {\bibfnamefont {P.}~\bibnamefont
  {Gaspard}},\ }\href@noop {} {\emph {\bibinfo {title} {{Chaos, Scattering and
  Statistical Mechanics}}}},\ \bibinfo {series} {Cambridge Nonlinear Science
  Series}, Vol.~\bibinfo {volume} {9}\ (\bibinfo  {publisher} {Cambridge
  University Press},\ \bibinfo {year} {1998})\BibitemShut {NoStop}%
\bibitem [{\citenamefont {Sz{\'a}sz}(2000)}]{Szasz:2000book}%
  \BibitemOpen
  \bibinfo {editor} {\bibfnamefont {D.}~\bibnamefont {Sz{\'a}sz}},\ ed.,\
  \href@noop {} {\emph {\bibinfo {title} {{Hard Ball Systems and the Lorentz
  Gas}}}},\ Encyclopaedia of Mathematical Sciences\ (\bibinfo  {publisher}
  {Springer},\ \bibinfo {year} {2000})\BibitemShut {NoStop}%
\bibitem [{\citenamefont {Cvitanovi{\'c}}\ \emph {et~al.}(2012)\citenamefont
  {Cvitanovi{\'c}}, \citenamefont {Artuso}, \citenamefont {Mainieri},
  \citenamefont {Tanner},\ and\ \citenamefont {Vattay}}]{Cvitanovic:2004p284}%
  \BibitemOpen
  \bibfield  {author} {\bibinfo {author} {\bibfnamefont {P.}~\bibnamefont
  {Cvitanovi{\'c}}}, \bibinfo {author} {\bibfnamefont {R.}~\bibnamefont
  {Artuso}}, \bibinfo {author} {\bibfnamefont {R.}~\bibnamefont {Mainieri}},
  \bibinfo {author} {\bibfnamefont {G.}~\bibnamefont {Tanner}}, \ and\ \bibinfo
  {author} {\bibfnamefont {G.}~\bibnamefont {Vattay}},\ }\href@noop {} {\emph
  {\bibinfo {title} {{Chaos: Classical and Quantum}}}}\ (\bibinfo  {publisher}
  {ChaosBook. org (Niels Bohr Institute, Copenhagen 2012)},\ \bibinfo {year}
  {2012})\BibitemShut {NoStop}%
\bibitem [{\citenamefont {Dettmann}(2000)}]{Dettmann:2000inSzasz}%
  \BibitemOpen
  \bibfield  {author} {\bibinfo {author} {\bibfnamefont {C.~P.}\ \bibnamefont
  {Dettmann}},\ }in\ \href@noop {} {\emph {\bibinfo {booktitle} {Hard Ball
  Systems and the Lorentz Gas}}},\ \bibinfo {editor} {edited by\ \bibinfo
  {editor} {\bibfnamefont {D.}~\bibnamefont {Sz{\'a}sz}}}\ (\bibinfo
  {publisher} {Springer},\ \bibinfo {year} {2000})\ pp.\ \bibinfo {pages}
  {315--365}\BibitemShut {NoStop}%
\bibitem [{\citenamefont {Gaspard}\ \emph {et~al.}(2003)\citenamefont
  {Gaspard}, \citenamefont {Nicolis},\ and\ \citenamefont
  {Dorfman}}]{Gaspard:2003p298}%
  \BibitemOpen
  \bibfield  {author} {\bibinfo {author} {\bibfnamefont {P.}~\bibnamefont
  {Gaspard}}, \bibinfo {author} {\bibfnamefont {G.}~\bibnamefont {Nicolis}}, \
  and\ \bibinfo {author} {\bibfnamefont {J.~R.}\ \bibnamefont {Dorfman}},\
  }\href@noop {} {\bibfield  {journal} {\bibinfo  {journal} {Physica A:
  Statistical Mechanics and its Applications}\ }\textbf {\bibinfo {volume}
  {323}},\ \bibinfo {pages} {294} (\bibinfo {year} {2003})}\BibitemShut
  {NoStop}%
\bibitem [{\citenamefont {Bunimovich}\ and\ \citenamefont
  {Sinai}(1981)}]{Bunimovich:1981p479}%
  \BibitemOpen
  \bibfield  {author} {\bibinfo {author} {\bibfnamefont {L.~A.}\ \bibnamefont
  {Bunimovich}}\ and\ \bibinfo {author} {\bibfnamefont {Y.~G.}\ \bibnamefont
  {Sinai}},\ }\href@noop {} {\bibfield  {journal} {\bibinfo  {journal}
  {Communications in Mathematical Physics}\ }\textbf {\bibinfo {volume} {78}},\
  \bibinfo {pages} {479} (\bibinfo {year} {1981})}\BibitemShut {NoStop}%
\bibitem [{\citenamefont {Bunimovich}\ \emph {et~al.}(1991)\citenamefont
  {Bunimovich}, \citenamefont {Sinai},\ and\ \citenamefont
  {Chernov}}]{Bunimovich:1991p47}%
  \BibitemOpen
  \bibfield  {author} {\bibinfo {author} {\bibfnamefont {L.~A.}\ \bibnamefont
  {Bunimovich}}, \bibinfo {author} {\bibfnamefont {Y.~G.}\ \bibnamefont
  {Sinai}}, \ and\ \bibinfo {author} {\bibfnamefont {N.}~\bibnamefont
  {Chernov}},\ }\href@noop {} {\bibfield  {journal} {\bibinfo  {journal}
  {Russian Mathematical Surveys}\ }\textbf {\bibinfo {volume} {46}},\ \bibinfo
  {pages} {47} (\bibinfo {year} {1991})}\BibitemShut {NoStop}%
\bibitem [{\citenamefont {Chernov}(2006)}]{Chernov:2006v122p1061}%
  \BibitemOpen
  \bibfield  {author} {\bibinfo {author} {\bibfnamefont {N.}~\bibnamefont
  {Chernov}},\ }\href@noop {} {\bibfield  {journal} {\bibinfo  {journal}
  {Journal of Statistical Physics}\ }\textbf {\bibinfo {volume} {122}},\
  \bibinfo {pages} {1061} (\bibinfo {year} {2006})}\BibitemShut {NoStop}%
\bibitem [{\citenamefont {Young}(1998)}]{Young:1998p136}%
  \BibitemOpen
  \bibfield  {author} {\bibinfo {author} {\bibfnamefont {L.-S.}\ \bibnamefont
  {Young}},\ }\href@noop {} {\bibfield  {journal} {\bibinfo  {journal} {Annals
  of Mathematics}\ }\textbf {\bibinfo {volume} {147}},\ \bibinfo {pages} {585}
  (\bibinfo {year} {1998})}\BibitemShut {NoStop}%
\bibitem [{\citenamefont {Schmidt}(1998)}]{Schmidt:1998v327p837}%
  \BibitemOpen
  \bibfield  {author} {\bibinfo {author} {\bibfnamefont {K.}~\bibnamefont
  {Schmidt}},\ }\href@noop {} {\bibfield  {journal} {\bibinfo  {journal}
  {Comptes Rendus de l'Acad{\'e}mie des Sciences - Series I - Mathematics}\
  }\textbf {\bibinfo {volume} {327}},\ \bibinfo {pages} {837} (\bibinfo {year}
  {1998})}\BibitemShut {NoStop}%
\bibitem [{\citenamefont {Conze}(1999)}]{Conze:1999v19p1233}%
  \BibitemOpen
  \bibfield  {author} {\bibinfo {author} {\bibfnamefont {J.-P.}\ \bibnamefont
  {Conze}},\ }\href@noop {} {\bibfield  {journal} {\bibinfo  {journal} {Ergodic
  Theory and Dynamical Systems}\ }\textbf {\bibinfo {volume} {19}},\ \bibinfo
  {pages} {1233} (\bibinfo {year} {1999})}\BibitemShut {NoStop}%
\bibitem [{\citenamefont {Lenci}(2003)}]{Lenci:2003v23p869}%
  \BibitemOpen
  \bibfield  {author} {\bibinfo {author} {\bibfnamefont {M.}~\bibnamefont
  {Lenci}},\ }\href@noop {} {\bibfield  {journal} {\bibinfo  {journal} {Ergodic
  Theory and Dynamical Systems}\ }\textbf {\bibinfo {volume} {23}},\ \bibinfo
  {pages} {869} (\bibinfo {year} {2003})}\BibitemShut {NoStop}%
\bibitem [{\citenamefont {Lenci}(2006)}]{Lenci:2006v26p799}%
  \BibitemOpen
  \bibfield  {author} {\bibinfo {author} {\bibfnamefont {M.}~\bibnamefont
  {Lenci}},\ }\href@noop {} {\bibfield  {journal} {\bibinfo  {journal} {Ergodic
  Theory and Dynamical Systems}\ }\textbf {\bibinfo {volume} {26}},\ \bibinfo
  {pages} {799} (\bibinfo {year} {2006})}\BibitemShut {NoStop}%
\bibitem [{\citenamefont {Cristadoro}\ \emph {et~al.}(2010)\citenamefont
  {Cristadoro}, \citenamefont {Lenci},\ and\ \citenamefont
  {Seri}}]{cristadoro:2010recurrence}%
  \BibitemOpen
  \bibfield  {author} {\bibinfo {author} {\bibfnamefont {G.}~\bibnamefont
  {Cristadoro}}, \bibinfo {author} {\bibfnamefont {M.}~\bibnamefont {Lenci}}, \
  and\ \bibinfo {author} {\bibfnamefont {M.}~\bibnamefont {Seri}},\ }\href@noop
  {} {\bibfield  {journal} {\bibinfo  {journal} {Chaos: An Interdisciplinary
  Journal of Nonlinear Science}\ }\textbf {\bibinfo {volume} {20}},\ \bibinfo
  {pages} {023115} (\bibinfo {year} {2010})}\BibitemShut {NoStop}%
\bibitem [{\citenamefont {Chernov}\ and\ \citenamefont
  {Dolgopyat}(2006)}]{Chernov:2006v2p1679}%
  \BibitemOpen
  \bibfield  {author} {\bibinfo {author} {\bibfnamefont {N.}~\bibnamefont
  {Chernov}}\ and\ \bibinfo {author} {\bibfnamefont {D.}~\bibnamefont
  {Dolgopyat}},\ }\href@noop {} {\bibfield  {journal} {\bibinfo  {journal}
  {Proceedings of the International Congress of Mathematicians: Madrid, August
  22-30, 2006: invited lectures}\ }\textbf {\bibinfo {volume} {2}},\ \bibinfo
  {pages} {1679} (\bibinfo {year} {2006})}\BibitemShut {NoStop}%
\bibitem [{\citenamefont {Dettmann}(2014)}]{dettmann:1402.7010}%
  \BibitemOpen
  \bibfield  {author} {\bibinfo {author} {\bibfnamefont {C.~P.}\ \bibnamefont
  {Dettmann}},\ }\href@noop {} {\bibfield  {journal} {\bibinfo  {journal}
  {arXiv preprint arXiv:1402.7010}\ } (\bibinfo {year} {2014})}\BibitemShut
  {NoStop}%
\bibitem [{\citenamefont {Bleher}(1992)}]{1992JSP....66..315B}%
  \BibitemOpen
  \bibfield  {author} {\bibinfo {author} {\bibfnamefont {P.~M.}\ \bibnamefont
  {Bleher}},\ }\href@noop {} {\bibfield  {journal} {\bibinfo  {journal}
  {Journal of Statistical Physics}\ }\textbf {\bibinfo {volume} {66}},\
  \bibinfo {pages} {315} (\bibinfo {year} {1992})}\BibitemShut {NoStop}%
\bibitem [{\citenamefont {Sanders}(2008)}]{Sanders:2008p453}%
  \BibitemOpen
  \bibfield  {author} {\bibinfo {author} {\bibfnamefont {D.~P.}\ \bibnamefont
  {Sanders}},\ }\href@noop {} {\bibfield  {journal} {\bibinfo  {journal}
  {Physical Review E}\ }\textbf {\bibinfo {volume} {78}},\ \bibinfo {pages}
  {060101} (\bibinfo {year} {2008})}\BibitemShut {NoStop}%
\bibitem [{\citenamefont {Dettmann}(2012)}]{Dettmann:2011p18216}%
  \BibitemOpen
  \bibfield  {author} {\bibinfo {author} {\bibfnamefont {C.~P.}\ \bibnamefont
  {Dettmann}},\ }\href@noop {} {\bibfield  {journal} {\bibinfo  {journal}
  {Journal of Statistical Physics}\ }\textbf {\bibinfo {volume} {146}},\
  \bibinfo {pages} {181} (\bibinfo {year} {2012})}\BibitemShut {NoStop}%
\bibitem [{\citenamefont {N{\'a}ndori}\ \emph {et~al.}(2012)\citenamefont
  {N{\'a}ndori}, \citenamefont {Sz{\'a}sz},\ and\ \citenamefont
  {Varj{\'u}}}]{Nandori:2012arXiv1210}%
  \BibitemOpen
  \bibfield  {author} {\bibinfo {author} {\bibfnamefont {P.}~\bibnamefont
  {N{\'a}ndori}}, \bibinfo {author} {\bibfnamefont {D.}~\bibnamefont
  {Sz{\'a}sz}}, \ and\ \bibinfo {author} {\bibfnamefont {T.}~\bibnamefont
  {Varj{\'u}}},\ }\href@noop {} {\bibfield  {journal} {\bibinfo  {journal}
  {arXiv}\ ,\ \bibinfo {pages} {1210.2231}} (\bibinfo {year}
  {2012})}\BibitemShut {NoStop}%
\bibitem [{\citenamefont {Zacherl}\ \emph {et~al.}(1986)\citenamefont
  {Zacherl}, \citenamefont {Geisel}, \citenamefont {Nierwetberg},\ and\
  \citenamefont {Radons}}]{Zacherl:1986p7768}%
  \BibitemOpen
  \bibfield  {author} {\bibinfo {author} {\bibfnamefont {A.}~\bibnamefont
  {Zacherl}}, \bibinfo {author} {\bibfnamefont {T.}~\bibnamefont {Geisel}},
  \bibinfo {author} {\bibfnamefont {J.}~\bibnamefont {Nierwetberg}}, \ and\
  \bibinfo {author} {\bibfnamefont {G.}~\bibnamefont {Radons}},\ }\href@noop {}
  {\bibfield  {journal} {\bibinfo  {journal} {Physics Letters A}\ }\textbf
  {\bibinfo {volume} {114}},\ \bibinfo {pages} {317} (\bibinfo {year}
  {1986})}\BibitemShut {NoStop}%
\bibitem [{\citenamefont {Friedman}\ and\ \citenamefont
  {Martin}(1984)}]{Friedman1984p23}%
  \BibitemOpen
  \bibfield  {author} {\bibinfo {author} {\bibfnamefont {B.}~\bibnamefont
  {Friedman}}\ and\ \bibinfo {author} {\bibfnamefont {R.~F.}\ \bibnamefont
  {Martin}, \bibfnamefont {Jr.}},\ }\href@noop {} {\bibfield  {journal}
  {\bibinfo  {journal} {Physics Letters A}\ }\textbf {\bibinfo {volume}
  {105}},\ \bibinfo {pages} {23} (\bibinfo {year} {1984})}\BibitemShut
  {NoStop}%
\bibitem [{\citenamefont {Friedman}\ and\ \citenamefont
  {Martin}(1988)}]{Friedman:1988v30p219}%
  \BibitemOpen
  \bibfield  {author} {\bibinfo {author} {\bibfnamefont {B.}~\bibnamefont
  {Friedman}}\ and\ \bibinfo {author} {\bibfnamefont {R.~F.}\ \bibnamefont
  {Martin}, \bibfnamefont {Jr.}},\ }\href@noop {} {\bibfield  {journal}
  {\bibinfo  {journal} {Physica D Nonlinear Phenomena}\ }\textbf {\bibinfo
  {volume} {30}},\ \bibinfo {pages} {219} (\bibinfo {year} {1988})}\BibitemShut
  {NoStop}%
\bibitem [{\citenamefont {Garrido}\ and\ \citenamefont
  {Gallavotti}(1994)}]{GarridoGallavottiBilliardCorrelationFnsJSP1994}%
  \BibitemOpen
  \bibfield  {author} {\bibinfo {author} {\bibfnamefont {P.~L.}\ \bibnamefont
  {Garrido}}\ and\ \bibinfo {author} {\bibfnamefont {G.}~\bibnamefont
  {Gallavotti}},\ }\href@noop {} {\bibfield  {journal} {\bibinfo  {journal}
  {Journal of Statistical Physics}\ }\textbf {\bibinfo {volume} {76}},\
  \bibinfo {pages} {549} (\bibinfo {year} {1994})}\BibitemShut {NoStop}%
\bibitem [{\citenamefont {Matsuoka}\ and\ \citenamefont
  {Martin}(1997)}]{Matsuoka:1997p776}%
  \BibitemOpen
  \bibfield  {author} {\bibinfo {author} {\bibfnamefont {H.}~\bibnamefont
  {Matsuoka}}\ and\ \bibinfo {author} {\bibfnamefont {R.~F.}\ \bibnamefont
  {Martin}, \bibfnamefont {Jr.}},\ }\href@noop {} {\bibfield  {journal}
  {\bibinfo  {journal} {Journal of Statistical Physics}\ }\textbf {\bibinfo
  {volume} {88}},\ \bibinfo {pages} {81} (\bibinfo {year} {1997})}\BibitemShut
  {NoStop}%
\bibitem [{\citenamefont {Dahlqvist}\ and\ \citenamefont
  {Artuso}(1996)}]{Dahlqvist:1996p16292}%
  \BibitemOpen
  \bibfield  {author} {\bibinfo {author} {\bibfnamefont {P.}~\bibnamefont
  {Dahlqvist}}\ and\ \bibinfo {author} {\bibfnamefont {R.}~\bibnamefont
  {Artuso}},\ }\href@noop {} {\bibfield  {journal} {\bibinfo  {journal}
  {Physics Letters A}\ }\textbf {\bibinfo {volume} {219}},\ \bibinfo {pages}
  {212} (\bibinfo {year} {1996})}\BibitemShut {NoStop}%
\bibitem [{\citenamefont {Melbourne}(2009)}]{Melbourne:2009v98p163}%
  \BibitemOpen
  \bibfield  {author} {\bibinfo {author} {\bibfnamefont {I.}~\bibnamefont
  {Melbourne}},\ }\href@noop {} {\bibfield  {journal} {\bibinfo  {journal}
  {Proceedings of the London Mathematical Society}\ }\textbf {\bibinfo {volume}
  {98}},\ \bibinfo {pages} {163} (\bibinfo {year} {2009})}\BibitemShut
  {NoStop}%
\bibitem [{\citenamefont {Sz{\'a}sz}\ and\ \citenamefont
  {Varj{\'u}}(2007)}]{Szasz:2007v129p59}%
  \BibitemOpen
  \bibfield  {author} {\bibinfo {author} {\bibfnamefont {D.}~\bibnamefont
  {Sz{\'a}sz}}\ and\ \bibinfo {author} {\bibfnamefont {T.}~\bibnamefont
  {Varj{\'u}}},\ }\href@noop {} {\bibfield  {journal} {\bibinfo  {journal}
  {Journal of Statistical Physics}\ }\textbf {\bibinfo {volume} {129}},\
  \bibinfo {pages} {59} (\bibinfo {year} {2007})}\BibitemShut {NoStop}%
\bibitem [{\citenamefont {B{\'a}lint}\ and\ \citenamefont
  {Gou{\"e}zel}(2006)}]{Balint:2006p18224}%
  \BibitemOpen
  \bibfield  {author} {\bibinfo {author} {\bibfnamefont {P.}~\bibnamefont
  {B{\'a}lint}}\ and\ \bibinfo {author} {\bibfnamefont {S.}~\bibnamefont
  {Gou{\"e}zel}},\ }\href@noop {} {\bibfield  {journal} {\bibinfo  {journal}
  {Communications in Mathematical Physics}\ }\textbf {\bibinfo {volume}
  {263}},\ \bibinfo {pages} {461} (\bibinfo {year} {2006})}\BibitemShut
  {NoStop}%
\bibitem [{\citenamefont {Dolgopyat}\ and\ \citenamefont
  {Chernov}(2009)}]{Dolgopyat:2009p16456}%
  \BibitemOpen
  \bibfield  {author} {\bibinfo {author} {\bibfnamefont {D.~I.}\ \bibnamefont
  {Dolgopyat}}\ and\ \bibinfo {author} {\bibfnamefont {N.~I.}\ \bibnamefont
  {Chernov}},\ }\href@noop {} {\bibfield  {journal} {\bibinfo  {journal}
  {Russian Mathematical Surveys}\ }\textbf {\bibinfo {volume} {64}},\ \bibinfo
  {pages} {651} (\bibinfo {year} {2009})}\BibitemShut {NoStop}%
\bibitem [{\citenamefont {Courbage}\ \emph {et~al.}(2008)\citenamefont
  {Courbage}, \citenamefont {Edelman}, \citenamefont {Fathi},\ and\
  \citenamefont {Zaslavsky}}]{Courbage:2008p454}%
  \BibitemOpen
  \bibfield  {author} {\bibinfo {author} {\bibfnamefont {M.}~\bibnamefont
  {Courbage}}, \bibinfo {author} {\bibfnamefont {M.}~\bibnamefont {Edelman}},
  \bibinfo {author} {\bibfnamefont {S.~M.~S.}\ \bibnamefont {Fathi}}, \ and\
  \bibinfo {author} {\bibfnamefont {G.~M.}\ \bibnamefont {Zaslavsky}},\
  }\href@noop {} {\bibfield  {journal} {\bibinfo  {journal} {Physical Review
  E}\ }\textbf {\bibinfo {volume} {77}},\ \bibinfo {pages} {036203} (\bibinfo
  {year} {2008})}\BibitemShut {NoStop}%
\bibitem [{\citenamefont {Dahlqvist}(1996)}]{Dahlqvist:1996p16294}%
  \BibitemOpen
  \bibfield  {author} {\bibinfo {author} {\bibfnamefont {P.}~\bibnamefont
  {Dahlqvist}},\ }\href@noop {} {\bibfield  {journal} {\bibinfo  {journal}
  {Journal of Statistical Physics}\ }\textbf {\bibinfo {volume} {84}},\
  \bibinfo {pages} {773} (\bibinfo {year} {1996})}\BibitemShut {NoStop}%
\bibitem [{\citenamefont {Sanders}\ and\ \citenamefont
  {Larralde}(2006)}]{Sanders:2006p452}%
  \BibitemOpen
  \bibfield  {author} {\bibinfo {author} {\bibfnamefont {D.~P.}\ \bibnamefont
  {Sanders}}\ and\ \bibinfo {author} {\bibfnamefont {H.}~\bibnamefont
  {Larralde}},\ }\href@noop {} {\bibfield  {journal} {\bibinfo  {journal}
  {Physical Review E}\ }\textbf {\bibinfo {volume} {73}},\ \bibinfo {pages}
  {026205} (\bibinfo {year} {2006})}\BibitemShut {NoStop}%
\bibitem [{\citenamefont {Chernov}(1997)}]{Chernov:1997p1}%
  \BibitemOpen
  \bibfield  {author} {\bibinfo {author} {\bibfnamefont {N.}~\bibnamefont
  {Chernov}},\ }\href@noop {} {\bibfield  {journal} {\bibinfo  {journal}
  {Journal of Statistical Physics}\ }\textbf {\bibinfo {volume} {88}},\
  \bibinfo {pages} {1} (\bibinfo {year} {1997})}\BibitemShut {NoStop}%
\bibitem [{\citenamefont {Armstead}\ \emph {et~al.}(2003)\citenamefont
  {Armstead}, \citenamefont {Hunt},\ and\ \citenamefont
  {Ott}}]{2003PhRvE..67b1110A}%
  \BibitemOpen
  \bibfield  {author} {\bibinfo {author} {\bibfnamefont {D.~N.}\ \bibnamefont
  {Armstead}}, \bibinfo {author} {\bibfnamefont {B.~R.}\ \bibnamefont {Hunt}},
  \ and\ \bibinfo {author} {\bibfnamefont {E.}~\bibnamefont {Ott}},\
  }\href@noop {} {\bibfield  {journal} {\bibinfo  {journal} {Physical Review
  E}\ }\textbf {\bibinfo {volume} {67}},\ \bibinfo {pages} {021110} (\bibinfo
  {year} {2003})}\BibitemShut {NoStop}%
\bibitem [{\citenamefont {Artuso}\ and\ \citenamefont
  {Cristadoro}(2003)}]{2003PhRvL..90x4101A}%
  \BibitemOpen
  \bibfield  {author} {\bibinfo {author} {\bibfnamefont {R.}~\bibnamefont
  {Artuso}}\ and\ \bibinfo {author} {\bibfnamefont {G.}~\bibnamefont
  {Cristadoro}},\ }\href@noop {} {\bibfield  {journal} {\bibinfo  {journal}
  {Physical Review Letters}\ }\textbf {\bibinfo {volume} {90}},\ \bibinfo
  {pages} {244101} (\bibinfo {year} {2003})}\BibitemShut {NoStop}%
\bibitem [{\citenamefont {Melbourne}\ and\ \citenamefont
  {T{\"o}r{\"o}k}(2011)}]{Melbounre:2012v32p1091}%
  \BibitemOpen
  \bibfield  {author} {\bibinfo {author} {\bibfnamefont {I.}~\bibnamefont
  {Melbourne}}\ and\ \bibinfo {author} {\bibfnamefont {A.}~\bibnamefont
  {T{\"o}r{\"o}k}},\ }\href@noop {} {\bibfield  {journal} {\bibinfo  {journal}
  {Ergodic Theory and Dynamical Systems}\ }\textbf {\bibinfo {volume} {32}},\
  \bibinfo {pages} {1091} (\bibinfo {year} {2011})}\BibitemShut {NoStop}%
\bibitem [{\citenamefont {Melbourne}(2012)}]{Melbourne:private}%
  \BibitemOpen
  \bibfield  {author} {\bibinfo {author} {\bibfnamefont {I.}~\bibnamefont
  {Melbourne}},\ }\href@noop {} {\bibfield  {journal} {\bibinfo  {journal}
  {Private communication}\ } (\bibinfo {year} {2012})}\BibitemShut {NoStop}%
\bibitem [{\citenamefont {Castiglione}\ \emph {et~al.}(1999)\citenamefont
  {Castiglione}, \citenamefont {Mazzino}, \citenamefont
  {Muratore-Ginanneschi},\ and\ \citenamefont
  {Vulpiani}}]{Castiglione:1999p690}%
  \BibitemOpen
  \bibfield  {author} {\bibinfo {author} {\bibfnamefont {P.}~\bibnamefont
  {Castiglione}}, \bibinfo {author} {\bibfnamefont {A.}~\bibnamefont
  {Mazzino}}, \bibinfo {author} {\bibfnamefont {P.}~\bibnamefont
  {Muratore-Ginanneschi}}, \ and\ \bibinfo {author} {\bibfnamefont
  {A.}~\bibnamefont {Vulpiani}},\ }\href@noop {} {\bibfield  {journal}
  {\bibinfo  {journal} {Physica D Nonlinear Phenomena}\ }\textbf {\bibinfo
  {volume} {134}},\ \bibinfo {pages} {75} (\bibinfo {year} {1999})}\BibitemShut
  {NoStop}%
\bibitem [{\citenamefont {Gnedenko}\ and\ \citenamefont
  {Kolmogorov}(1968)}]{gnedenko1968limit}%
  \BibitemOpen
  \bibfield  {author} {\bibinfo {author} {\bibfnamefont {B.~V.}\ \bibnamefont
  {Gnedenko}}\ and\ \bibinfo {author} {\bibfnamefont {A.~N.}\ \bibnamefont
  {Kolmogorov}},\ }\href@noop {} {\emph {\bibinfo {title} {{Limit Distributions
  for Sums of Independent Random Variables}}}},\ \bibinfo {edition} {revised}\
  ed.,\ Addison-Wesley series in statistics\ (\bibinfo  {publisher}
  {Addison-Wesley},\ \bibinfo {year} {1968})\BibitemShut {NoStop}%
\bibitem [{\citenamefont {Knight}(2000)}]{Knight:2000MathStat}%
  \BibitemOpen
  \bibfield  {author} {\bibinfo {author} {\bibfnamefont {K.}~\bibnamefont
  {Knight}},\ }\href@noop {} {\emph {\bibinfo {title} {{Mathematical
  Statistics}}}}\ (\bibinfo  {publisher} {Chapman and Hall/CRC},\ \bibinfo
  {year} {2000})\BibitemShut {NoStop}%
\bibitem [{\citenamefont {Chernov}\ \emph {et~al.}(2012)\citenamefont
  {Chernov}, \citenamefont {B\'alint},\ and\ \citenamefont
  {Dolgopyat}}]{Chernovetal:private}%
  \BibitemOpen
  \bibfield  {author} {\bibinfo {author} {\bibfnamefont {N.}~\bibnamefont
  {Chernov}}, \bibinfo {author} {\bibfnamefont {P.}~\bibnamefont {B\'alint}}, \
  and\ \bibinfo {author} {\bibfnamefont {D.}~\bibnamefont {Dolgopyat}},\
  }\href@noop {} {\bibfield  {journal} {\bibinfo  {journal} {Private
  communication}\ } (\bibinfo {year} {2012})}\BibitemShut {NoStop}%
\bibitem [{\citenamefont {B{\'a}lint}\ \emph {et~al.}(2011)\citenamefont
  {B{\'a}lint}, \citenamefont {Chernov},\ and\ \citenamefont
  {Dolgopyat}}]{Balint:2011p18227}%
  \BibitemOpen
  \bibfield  {author} {\bibinfo {author} {\bibfnamefont {P.}~\bibnamefont
  {B{\'a}lint}}, \bibinfo {author} {\bibfnamefont {N.}~\bibnamefont {Chernov}},
  \ and\ \bibinfo {author} {\bibfnamefont {D.}~\bibnamefont {Dolgopyat}},\
  }\href@noop {} {\bibfield  {journal} {\bibinfo  {journal} {Communications in
  Mathematical Physics}\ }\textbf {\bibinfo {volume} {308}},\ \bibinfo {pages}
  {479} (\bibinfo {year} {2011})}\BibitemShut {NoStop}%
\bibitem [{Note1()}]{Note1}%
  \BibitemOpen
  \bibinfo {note} {In other words, if $x = (\mathbf{r},\mathbf{v})$
    denotes the phase-space coordinates at a grazing collision 
    point $\mathbf {r}$, i.e.~such that $\mathbf{v}$ is
    tangent to the scatterer, $x$ is mapped to a point $x_1=
    (\mathbf{r}_1, \mathbf{v})$ by the collision map whose velocity component
    remains unchanged. The vector connecting the two successive positions is
    $\boldsymbol{\psi}(x) = \mathbf{r}_1 - \mathbf{r}$.}\BibitemShut {Stop}%
\bibitem [{\citenamefont {Zaslavsky}\ and\ \citenamefont
  {Edelman}(1997)}]{Zaslavsky:1997p706}%
  \BibitemOpen
  \bibfield  {author} {\bibinfo {author} {\bibfnamefont {G.~M.}\ \bibnamefont
  {Zaslavsky}}\ and\ \bibinfo {author} {\bibfnamefont {M.}~\bibnamefont
  {Edelman}},\ }\href@noop {} {\bibfield  {journal} {\bibinfo  {journal}
  {Physical Review E}\ }\textbf {\bibinfo {volume} {56}},\ \bibinfo {pages}
  {5310} (\bibinfo {year} {1997})}\BibitemShut {NoStop}%
\bibitem [{Note2()}]{Note2}%
  \BibitemOpen
  \bibinfo {note} {We are thereby discarding the possibility that other
  time-dependent terms diverging slower than the logarithm, such as, for
  instance, $\protect \qopname \relax o{log}\protect \qopname \relax o{log}t$,
  may be present; we do not find numerical evidence to support the existence of
  such terms.}\BibitemShut {Stop}%
\bibitem [{\citenamefont {Cristadoro}\ \emph
  {et~al.}({\natexlab{a}})\citenamefont {Cristadoro}, \citenamefont {Gilbert},
  \citenamefont {Lenci},\ and\ \citenamefont {Sanders}}]{us:short}%
  \BibitemOpen
  \bibfield  {author} {\bibinfo {author} {\bibfnamefont {G.}~\bibnamefont
  {Cristadoro}}, \bibinfo {author} {\bibfnamefont {T.}~\bibnamefont {Gilbert}},
  \bibinfo {author} {\bibfnamefont {M.}~\bibnamefont {Lenci}}, \ and\ \bibinfo
  {author} {\bibfnamefont {D.~P.}\ \bibnamefont {Sanders}},\ }\href@noop {}
  {\bibfield  {journal} {\bibinfo  {journal} {unpublished}\ }
  ({\natexlab{a}})}\BibitemShut {NoStop}%
\bibitem [{\citenamefont {Cristadoro}\ \emph
  {et~al.}({\natexlab{b}})\citenamefont {Cristadoro}, \citenamefont {Gilbert},
  \citenamefont {Lenci},\ and\ \citenamefont {Sanders}}]{us:long}%
  \BibitemOpen
  \bibfield  {author} {\bibinfo {author} {\bibfnamefont {G.}~\bibnamefont
  {Cristadoro}}, \bibinfo {author} {\bibfnamefont {T.}~\bibnamefont {Gilbert}},
  \bibinfo {author} {\bibfnamefont {M.}~\bibnamefont {Lenci}}, \ and\ \bibinfo
  {author} {\bibfnamefont {D.~P.}\ \bibnamefont {Sanders}},\ }\href@noop {}
  {\bibfield  {journal} {\bibinfo  {journal} {unpublished}\ }
  ({\natexlab{b}})}\BibitemShut {NoStop}%
\bibitem [{Note3()}]{Note3}%
  \BibitemOpen
  \bibinfo {note} {This statement applies to type $(0,1)$ corridors. The
  frequency of occurrence of corridors of other types changes according to
  their relative widths.}\BibitemShut {Stop}%
\bibitem [{\citenamefont {Lubachevsky}(1991)}]{Lubachevsky:y1991v94p255}%
  \BibitemOpen
  \bibfield  {author} {\bibinfo {author} {\bibfnamefont {B.~D.}\ \bibnamefont
  {Lubachevsky}},\ }\href@noop {} {\bibfield  {journal} {\bibinfo  {journal}
  {Journal of Computational Physics}\ }\textbf {\bibinfo {volume} {94}},\
  \bibinfo {pages} {255} (\bibinfo {year} {1991})}\BibitemShut {NoStop}%
\bibitem [{\citenamefont {Hsu}\ and\ \citenamefont
  {Grassberger}(2011)}]{hsu:2011review}%
  \BibitemOpen
  \bibfield  {author} {\bibinfo {author} {\bibfnamefont {H.-P.}\ \bibnamefont
  {Hsu}}\ and\ \bibinfo {author} {\bibfnamefont {P.}~\bibnamefont
  {Grassberger}},\ }\href@noop {} {\bibfield  {journal} {\bibinfo  {journal}
  {Journal of Statistical Physics}\ }\textbf {\bibinfo {volume} {144}},\
  \bibinfo {pages} {597} (\bibinfo {year} {2011})}\BibitemShut {NoStop}%
\bibitem [{\citenamefont {Laffargue}\ and\ \citenamefont
  {Tailleur}(2014)}]{laffargue:2014locating}%
  \BibitemOpen
  \bibfield  {author} {\bibinfo {author} {\bibfnamefont {T.}~\bibnamefont
  {Laffargue}}\ and\ \bibinfo {author} {\bibfnamefont {J.}~\bibnamefont
  {Tailleur}},\ }\href@noop {} {\bibfield  {journal} {\bibinfo  {journal}
  {arXiv preprint arXiv:1404.2600}\ } (\bibinfo {year} {2014})}\BibitemShut
  {NoStop}%
\bibitem [{\citenamefont {Leit{\~a}o}\ \emph {et~al.}(2013)\citenamefont
  {Leit{\~a}o}, \citenamefont {Lopes},\ and\ \citenamefont
  {Altmann}}]{leitao:2013monte}%
  \BibitemOpen
  \bibfield  {author} {\bibinfo {author} {\bibfnamefont {J.~C.}\ \bibnamefont
  {Leit{\~a}o}}, \bibinfo {author} {\bibfnamefont {J.~V.~P.}\ \bibnamefont
  {Lopes}}, \ and\ \bibinfo {author} {\bibfnamefont {E.~G.}\ \bibnamefont
  {Altmann}},\ }\href@noop {} {\bibfield  {journal} {\bibinfo  {journal}
  {Physical review letters}\ }\textbf {\bibinfo {volume} {110}},\ \bibinfo
  {pages} {220601} (\bibinfo {year} {2013})}\BibitemShut {NoStop}%
\bibitem [{\citenamefont {Viswanathan}\ \emph {et~al.}(2011)\citenamefont
  {Viswanathan}, \citenamefont {da~Luz}, \citenamefont {Raposo},\ and\
  \citenamefont {Stanley}}]{viswanathan:2011physics}%
  \BibitemOpen
  \bibfield  {author} {\bibinfo {author} {\bibfnamefont {G.~M.}\ \bibnamefont
  {Viswanathan}}, \bibinfo {author} {\bibfnamefont {M.~G.~E.}\ \bibnamefont
  {da~Luz}}, \bibinfo {author} {\bibfnamefont {E.~P.}\ \bibnamefont {Raposo}},
  \ and\ \bibinfo {author} {\bibfnamefont {H.~E.}\ \bibnamefont {Stanley}},\
  }\href@noop {} {\emph {\bibinfo {title} {The Physics of Foraging: an
  Introduction to Random Searches and Biological Encounters}}}\ (\bibinfo
  {publisher} {Cambridge University Press},\ \bibinfo {address} {Cambrdige,
  UK},\ \bibinfo {year} {2011})\BibitemShut {NoStop}%
\bibitem [{\citenamefont {Geisel}\ \emph {et~al.}(1985)\citenamefont {Geisel},
  \citenamefont {Nierwetberg},\ and\ \citenamefont
  {Zacherl}}]{Geisel:1985p8023}%
  \BibitemOpen
  \bibfield  {author} {\bibinfo {author} {\bibfnamefont {T.}~\bibnamefont
  {Geisel}}, \bibinfo {author} {\bibfnamefont {J.}~\bibnamefont {Nierwetberg}},
  \ and\ \bibinfo {author} {\bibfnamefont {A.}~\bibnamefont {Zacherl}},\
  }\href@noop {} {\bibfield  {journal} {\bibinfo  {journal} {Physical Review
  Letters}\ }\textbf {\bibinfo {volume} {54}},\ \bibinfo {pages} {616}
  (\bibinfo {year} {1985})}\BibitemShut {NoStop}%
\bibitem [{\citenamefont {Zumofen}\ and\ \citenamefont
  {Klafter}(1993)}]{Zumofen:1993p804}%
  \BibitemOpen
  \bibfield  {author} {\bibinfo {author} {\bibfnamefont {G.}~\bibnamefont
  {Zumofen}}\ and\ \bibinfo {author} {\bibfnamefont {J.}~\bibnamefont
  {Klafter}},\ }\href@noop {} {\bibfield  {journal} {\bibinfo  {journal}
  {Physical Review E}\ }\textbf {\bibinfo {volume} {47}},\ \bibinfo {pages}
  {851} (\bibinfo {year} {1993})}\BibitemShut {NoStop}%
\bibitem [{\citenamefont {Machta}\ and\ \citenamefont
  {Zwanzig}(1983)}]{Machta:1983p182}%
  \BibitemOpen
  \bibfield  {author} {\bibinfo {author} {\bibfnamefont {J.}~\bibnamefont
  {Machta}}\ and\ \bibinfo {author} {\bibfnamefont {R.}~\bibnamefont
  {Zwanzig}},\ }\href@noop {} {\bibfield  {journal} {\bibinfo  {journal}
  {Physical Review Letters}\ }\textbf {\bibinfo {volume} {50}},\ \bibinfo
  {pages} {1959} (\bibinfo {year} {1983})}\BibitemShut {NoStop}%
\end{thebibliography}

%

\end{document}